\documentclass[useAMS,usenatbib]{mn2e}
\usepackage{graphicx}
\usepackage[euler]{textgreek}
\usepackage{breqn}
\usepackage{amssymb}

\title[Magnetic reconnection: a CR accelerator]{A magnetic reconnection model for explaining the multi-wavelength emission of the microquasars Cyg X-1 and Cyg X-3}
\author[B. Khiali, E. M. de Gouveia Dal Pino, M. V. del Valle]{B. Khiali$^{1}$\thanks{E-mail:
bkhiali@usp.br}, E. M. de Gouveia Dal Pino$^{1}$ and M. V. del Valle$^{2}$ \\
$^{1}$IAG-Universidade de S\~ao Paulo, Rua do Mat\~ao 1226, S\~ao Paulo, SP, Brazil\\
$^{2}$Instituto Argentino de Radioastronomia (IAR), CCT La Plata, C.C.5, 1894 Villa Elisa, Buenos Aires, Argentina}

\begin{document}


\pagerange{\pageref{firstpage}--\pageref{lastpage}} \pubyear{2014}

\maketitle

\label{firstpage}

\begin{abstract}
Recent studies have indicated that cosmic ray acceleration by a first-order Fermi process in magnetic reconnection current sheets can be efficient enough in the surrounds of compact sources. 
In this work, we discuss  this acceleration mechanism operating in the core region of galactic black hole binaries (or microquasars) and show the conditions under which this can be more efficient than shock acceleration. In addition, 
we compare the corresponding acceleration rate with the relevant radiative loss rates obtaining the possible energy cut-off of the accelerated particles and also compute the expected spectral energy distribution (SED) for two sources of this class, namely  Cygnus X-1 and  Cygnus X-3, considering both leptonic and hadronic processes.  The derived SEDs are comparable to the observed ones in the low and high energy ranges. Our results suggest that hadronic non-thermal emission due to photo-meson production may produce the very high energy gamma-rays in these microquasars.
\end{abstract}


\section{Introduction}

Detected non-thermal  radio to gamma-ray emission from  galactic binary systems hosting  stellar mass black holes, also denominated black hole binaries (BHBs), microquasars, or simply $\mu$QSOs \citep{MirableRodriquez1994,Tingay1995,HjellmingRupen1995}, provide clear evidence of the production of relativistic particles in their jets and probably also in the  innermost regions very close to the black hole (BH).  Currently, more than a dozen microquasars have been detected in the galaxy (\citealt{zhang13}). 

Generally, these sources are far from being stable and individual systems have often complex emission structure. Nevertheless,  all classes of BHBs exhibit common features and show basically two major states when considering their X-ray emission (2-100 keV): a quiescent and an outburst state (e.g., \citealt{remillard06}). The former is characterized by low X-ray luminosities and hard non-thermal spectra. Usually, transient BHBs exhibit this state for long periods, which allows one to obtain typical physical parameters of the system. On the other hand, the outburst state corresponds to intense activity and emission, and can be sub-classified  in three main active and many intermediary states. According to \citealt{remillard06} (see also \citealt{zhang13}), the three main active states are the thermal state (TS), the hard state (HS) and the steep power law state (SPLS). These states are usually explained as changes in the structure of the accretion flow, as remarked before. During the TS,  the soft X-ray thermal emission is believed to come from the inner region of the thin accretion disk that extends until the last stable orbits around the black hole. On the other hand, during the HS the observed weak thermal component suggests that the disk has been  truncated at a few hundreds/thousands gravitational radii. The hard X-ray emission  measured during this state is dominated by a power-law (PL) component and is often attributed to inverse Compton scattering of soft photons from the outer disk by relativistic electrons in the hot inner region of the system (e.g., \citealt{remillard06, malzak07}). The SPLS is almost a combination of the above two states, but the PL is steeper.

The observed  radio  and infra-red (IR) emission in microquasars is normally interpreted as due to synchrotron radiation produced by  relativistic particles in the jet outflow.

More recently a few microquasars  have been also detected in the gamma-ray range with \textit{AGILE} \citep{Tavanietal.2009,Bulgarelli2010,Sabatini2010a,Sabatini2010b,Sabatini2013}, \textit{Fermi-LAT} \citep{b4,b12} and \textit{MAGIC} \citep{Lorentz 2004}.  For Cygnus X-1 (Cyg X-1), for instance,  upper limits with 95\% confidence level have been obtained in the range of $\geq 150$ GeV \citep{albert2007}, while  in the case of Cygnus X-3 (Cyg X-3), upper limits of integrated gamma-ray flux above 250 GeV have been inferred by \cite{aleksic10}. Upper limits in the 0.1-10 GeV range have been also suggested for  GRS 1915+105 and GX 339-4.

There is no definite mechanism yet to explain the origin of the very high energy  (VHE) emission in microquasars. The main reason for this  is that the current sensitivity of the gamma-ray instruments is too poor to establish  the location of this emission in the source (e.g., \citealt{b12}).

 Regardless of the uncertainties, several models have been proposed, especially for Cyg X-1 and Cyg X-3.   \citealt{romero03}, for instance,  assumed that the gamma-ray emission is produced in a hadronic jet as a result of  the decay of neutral pions created in photon-ion collisions. 
An alternative model developed  by \citealt{BoschRamonetal.(2005b)}  assumed  that relativistic  protons also produced  in the jet may diffuse  through the interstellar medium (ISM) and then interact with molecular clouds and produce  gamma-rays out of $pp$ interactions via neutral pion decay.
Another model has been proposed by \citealt{Pianoet2012} in which both, leptonic (via inverse Compton) and hadronic (via neutral pion decay) might account for the observed gamma-ray emission.

All models above postulate that the primary relativistic particles (electrons and protons) are produced  behind shocks in the jet outflow. 

An alternative  mechanism  has been explored first in the context of microquasars (\citealt{gl05}, hereafter GL05) and later extended  to the framework of active galactic nuclei (AGNs) (\citealt{beteluis2010a}, hereafter GPK10) in which particles are accelerated in the surrounds of the BH of these sources, near the jet basis, by a first-order Fermi process, as proposed in GL05, within magnetic reconnection current sheets produced in  $fast$ encounters of the field lines arising  from the accretion disk and those of the BH magnestosphere.

Fast magnetic reconnection, which occurs when two magnetic fluxes of opposite polarity encounter each other and partially annihilate  very efficiently at a speed $V_R$  of the
order of the local Alfv\'en speed ($V_{A}$),   has been detected in  laboratory plasma experiments (e.g., \citealt{yamada_etal_10}) as well as in space  environments, like the earth magnetotail and the solar corona (see e.g., \citealt{deng01,su13}). Extensive  numerical work has been also carried out considering  collisionless (e.g., \citealt{zenitani01,zenitani09,drake06,drake10,cerutti13,cerutti14,sironi14}) and collisional flows (e.g., \citealt{kowal09,kowal2012,shibata01,loureiro07}). 
Different processes such as  kinetic plasma instabilities \citep{shay_etal_98, shay_etal_04, yamada_etal_10}, anomalous resistivity (e.g., \citealt{parker79,biskamp97,shay_etal_04}),  or turbulence (\citealt{LV99}, hereafter LV99, \citealt{kowal09, eyink11}), can lead to fast reconnection. The latter process in particular,  has been found  to be very efficient  because it provokes the wandering of the magnetic field lines allowing for several simultaneous events of reconnection within the current sheet (see \S 2.2).  

Fast reconnection has recently gained increasing interest also in other astrophysical contexts beyond the solar system because of its potential efficiency to explain magnetic field diffusion, dynamo process, and particle acceleration in different classes of  sources and environments  - from compact objects, like BHs (e.g., GL05, GL10, \citealt{Giannios2010}), pulsars (e.g., \citealt{cerutti13,cerutti14,sironi14}), and gamma ray bursts (e.g., \citealt{zhang11}), to more diffuse regions like the interstellar medium (ISM), intergalactic medium (IGM),  and star forming regions (e.g., \citealt{santoslima10, santoslima12,santoslima13,leao13};  see also  \citealt{uzdensky11,bete13,bete14} and references therein for reviews).

In the mechanism proposed by GL05,  particles are accelerated to relativistic velocities within the 
fast magnetic reconnection sheet in a similar way to the first-order Fermi process that occurs
in shocks, i.e.,  trapped charged particles may bounce back
and forth several times and gain energy due to head-on collisions with the two
converging magnetic fluxes of opposite polarity (see \S 2.3).
This acceleration mechanism has been also successfully tested numerically both in collisionless  by means of two-dimensional (2D)  PIC simulations (e.g.,  \citealt{drake06,zenitani01,zenitani09,drake10,cerutti13,cerutti14,sironi14}) and collisional magnetic reconnection sheets by means of 2D and 3D MHD simulations with test particles (\citealt{kowal2011,kowal2012}). 
Furthermore,  this  process  has been explored in depth in the natural laboratories of fast reconnection provided by solar flares \cite[e.g.,][]{drake06,lazarian09,drake10,gordovskyy10,gordovskyy11,zharkova11} and the earth magnetotail. For instance,  \cite{lazarian09} verified that the anomalous cosmic rays measured by Voyager  seem to be indeed accelerated in the reconnection regions of the magnetopause \cite[see also][]{drake10}.  In another study, \cite{lazarian10} invoked   the same mechanism to explain the excess of cosmic rays in the sub-TeV and multi-TeV ranges in the wake produced as the Solar system moves through interstellar gas.  
Magnetic reconnection has been also invoked in the production of ultra high energy cosmic rays  (e.g.,   \citealt{koteraolinto11}) and in particle acceleration in  astrophysical jets and gamma-ray bursts\citep{Giannios2010,delvalle11,zhang11}.

In the context of BHs, GPK10  found that the energy power extracted from events of fast magnetic reconnection between the magnetosphere of the BH  and the lines rising from the inner accretion disk can be more than sufficient to accelerate primary particles and produce the observed core radio synchrotron  radiation from microquasars and low luminosity AGNs (LLAGNs). Moreover,  they proposed that the observed correlation between the radio emission and the BH mass of these sources, spanning $10^{10}$ orders of magnitude in mass (in the so called fundamental plane of BHs, \citealt{merloni}), might be related to this process.
More recently,  Kadowaki, de Gouveia Dal Pino \& Singh 2014 (\citealt{beteluis2014}, henceforth KGS14) revisited this model exploring different mechanisms of fast magnetic reconnection and extended the study to include also the gamma-ray emission of a much larger sample  containing over  two hundred sources. They found that both LLAGNs and microquasars confirm the earlier trend found by GL05 and GPK10. Furthermore, when driven by turbulence, not only the radio but also the  gamma-ray emission  of these sources can be due to the magnetic power released by fast reconnection allowing for particle acceleration to relativistic velocities in the core region of these sources.
In another concomitant work \citealt{singh14} (hereafter SGK14), have repeated the analysis above of KGS14, but instead of employing the standard accretion disk/coronal model to describe the BH surrounds, they adopted an MADAF (magnetically advected accretion flow) and obtained very similar results to those of KGS14, 
for the same large sample of LLAGNs and microquasars.

In addition, it has been argued  in these studies that the fast  magnetic reconnection events could be directly related to the transition between  the hard  and the soft steep-power-law (SPLS) X-ray states seen in  microquasars, as described above.

Lately, similar mechanisms involving magnetic activity, reconnection and acceleration in the core regions of compact sources to explain their emission spectra have been also invoked by other works (e. g., \citealt{lyubarsky08,igumenshchev09,soker_10,uzdensky_spitkovsky_14,huang_etal_14}). In particular, magnetic reconnection between the magnetospheric lines of the central source and those anchored into the accretion disk resulting in the ejection of plasmons has been detected in numerical MHD studies by \citep[see, e.g.,][]{romanova_etal_02, romanova_etal_11, zanni_ferreira_09, zanni_ferreira_13, cemeljic_etal_13}. 
The recent numerical relativistic MHD simulations of magnetically arrested accretion disks by \citealt{tchekhovskoy11,mckinney12,dexter_etal_14} also evidence the development of magnetic reconnection in the magnetosphere of the BH and are consistent with our scenario above.

The  results above, and specially the correlations found in the works of KGS14 and SGK14 between the magnetic reconnection power released by turbulent driven fast reconnection in the surrounds of BHs  and the observed core radio and gamma-ray emission of a sample containing more than 200 sources  of microquasars and LLAGNS  (see figures 7 in KGS14 and 3 in SGK14), have  motivated  us to perform the present study,  undertaking a detailed multifrequency analysis of the non-thermal emission  of two well investigated observationally  microquasars, namely  Cyg X-1 and Cyg X-3 (which are also in the KGS14 and SGK14 samples), 
 aiming at reproducing their observed spectral energy distribution (SED) from radio to gamma-rays during outburst states. As in GL05, GPK10, KGS14 and SGK14,  we explore the potential effects of the interactions between the magnetosphere of the BH and the magnetic field lines that rise from  the  accretion disk. These magnetic fields are considered essential ingredients in most accretion disk/BH models to help to explain the variety and complexity of observed data (e.g., \citealt{zhang13,neronov07}), but are, in general, paradoxically neglected or avoided in the discussion of the acceleration and emission mechanisms in the nuclear regions of these compact sources. 
 
 We here compute the power released by fast magnetic reconnection between these two magnetic fluxes and then the resulting particle spectrum of accelerated particles in the magnetic reconnection site. In particular, we explore the first-order Fermi  acceleration process that may occur within the current sheet as  proposed in GL05. 

  We finally consider the relevant radiative loss mechanisms due to the interactions of the accelerated particles with the ambient matter, magnetic and radiation fields, and also  assess the importance of the acceleration by magnetic reconnection in comparison to  shock acceleration.

The outline of the paper is as follows. In Section 2, we describe in detail our acceleration model, while the equations employed to calculate the emission processes from radio to gamma-ray energies are presented in section 3. In Sections 4 and 5, we show the results of the application of the acceleration and emission model to Cyg X-1 and Cyg X-3, respectively. Finally in Section 6, we summarize our results and draw our conclusions. 

 \begin{figure}
  \centering
  \includegraphics[width=0.50\textwidth]{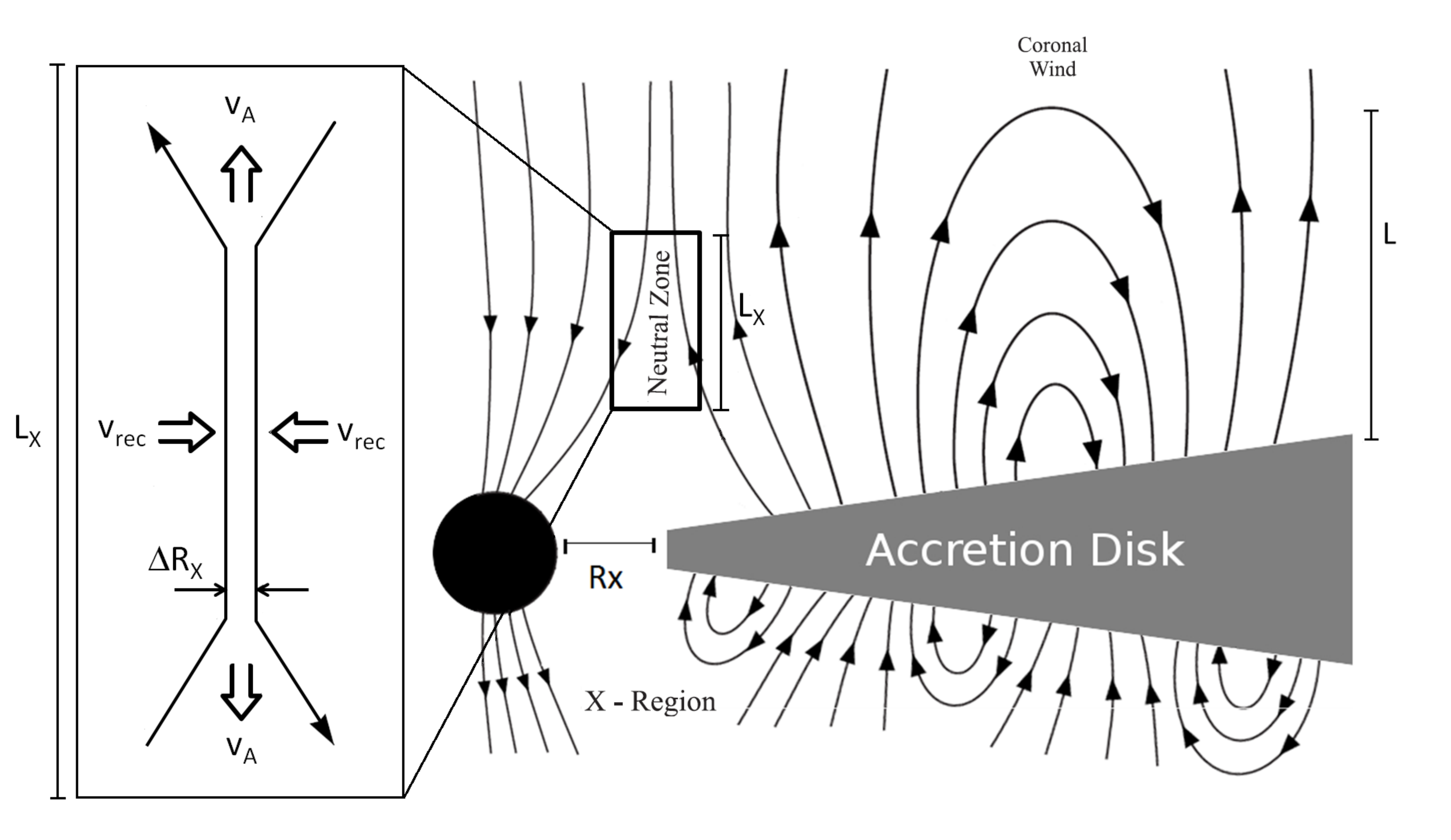}
  \caption{Scheme of magnetic reconnection between the lines rising from the accretion disk and the lines anchored into the BH horizon. Particle acceleration may occur in the magnetic reconnection site (neutral zone) by a first order Fermi process (adapted from GL05). }
  \label{wide_fig}
 \end{figure}


\section{Our Particle Acceleration Scenario}
We assume here that  relativistic particles may be accelerated in the core of the microquasar in the surrounds of the BH, near the basis of the jet launching region, as a result of events of  fast magnetic reconnection.
As stressed, this acceleration model  has been described in detail in GL05 and GPK10 and recently revisited in KGS14. We summarize here its main assumptions. As  in these former studies, we assume that the inner region of the accretion disk/corona system  alternates between two states which are controlled by changes in the global magnetic field. Right before a fast magnetic reconnection event, we assume that the system is in a state that possibly characterizes the transition from the hard to the soft state  as described in the previous section, and adopt a magnetized  accretion disk with a corona around the BH.

\subsection{The accretion disk/coronal fluid around the BH}
Although there is still much speculation on what should be the strength and geometry of the magnetic fields in the surrounds of  BHs, these are  necessary ingredients in order to explain, e.g.,  the formation of  narrow relativistic jets.
We consider here a scenario with the simplest possible configuration  
by considering a magnetized standard (geometrically thin and optically thick) accretion disk around the BH as in the cartoon of Fig. 1. 

A magnetosphere around the central BH can  be established from the drag of magnetic field lines by the accretion disk.  The  large-scale poloidal magnetic field in the disk corona can in turn be formed by the action of a turbulent dynamo inside the accretion disk (see GL05, KGL14 and references therein) or dragged from the surroundings.  This poloidal magnetic flux under  the action of the  disk differential rotation gives rise to a wind that partially removes angular momentum from the system and increases the accretion rate.
This  also increases the ram pressure of the accreting material that will then press the magnetic lines in the inner disk region against the lines anchored into the BH horizon allowing them to reconnect fast (see Figure 1). Momentum flux conservation between the magnetic pressure of the BH magnetosphere and the accreting flux determines the magnetic field intensity in this inner  region.

\subsection{Conditions for fast reconnection in the surrounds of the BH}

As discussed in \S 1 (see also GL05, GPK10, and KGS14), in the presence of kinetic plasma instabilities \citep{shay_etal_98, shay_etal_04, yamada_etal_10}, anomalous resistivity (e.g., \citealt{parker79,biskamp97,shay_etal_04}),  or turbulence (\citealt{LV99,kowal09, kowal2012}), reconnection may become very efficient and fast. 

The strongly magnetized and low dense coronal fluid of the systems we are dealing with in this work satisfies the condition  $L> l_{mfp} > r_l$ (where $L$ is the typical large scale dimension of the system, $l_{mfp}$ the ion mean free path and $r_l$ the ion Larmor radius). For such flows a weakly collisional or effectively collisional MHD description is more than appropriate (e.g. \citealt{liu03}) and  we will employ this approach here, as in GL05, GPK10, and KGS14\footnote{We should further notice  that  the BH of these systems is surrounded by accreting flow from the stellar companion which also favours a nearly collisional MHD approach.} 

In these MHD flows, a collisional turbulent fast reconnection approach is expected to be dominant (see KGL14).
According to the LV99 model, the presence of even weak turbulence causes the wandering of the magnetic field lines which allows for many independent patches to reconnect simultaneously making the global reconnection rate  large, $V_R \sim v_A (l_{inj}/L)^{1/2} (v_{turb}/v_A)^2$, where $V_R$ is the reconnection speed, and $l_{inj}$ and $v_{turb}$ the injection scale and velocity of the turbulence, respectively. This expression indicates that the reconnection rate can be as large as $\sim V_A$, which in the systems here considered may be near the light speed (see also KGL14). 
This theory has been extensively investigated (e.g.  \citealt{eyink11,lazarian12}) and  confirmed numerically by means of 3D MHD simulations (\citealt{kowal09,kowal2012}). In particular, it has been shown  (\citealt{eyink11}) that turbulent collisional fast reconnection prevails when the thickness of the current sheet (see eq. 4 below) is larger than the ion Larmor radius. As demonstrated in KGS14,  for the systems we are studying this condition is naturally satisfied and we will adopt this model to derive the magnetic power released by fast reconnection. \footnote{It should be noticed that GL05, GPK10 and KGS14 have also investigated another mechanism to induce fast magnetic reconnection based on anomalous resistivity (AR). This occurs in the presence of current driven instabilities that can enhance the microscopic Ohmic resistivity and speed up reconnection to rates much larger than that probed by the latter. On the other hand,  AR results rates which are much smaller than reconnection driven by turbulence as it prevails only at very small scales of the fluid. In fact, as shown in KGL14, AR  predicts a much thinner reconnection region and is unable to reproduce the observed emission for most of the sources investigated. In particular, in the case of  Cyg X-1 and Cyg X-3, the magnetic power released by fast reconnection driven by AR cannot accelerate particles to energies larger than $10^{12}$ eV (see more details in KGS14). Other instabilities, like e.g., tearing mode or Hall effect are also relevant to drive fast reconnection but only at very small scales as well, and are thus more appropriate for collisionless fluids (see \citealt{eyink11}).
}

The employment of a fast magnetic reconnection model driven by turbulence as in LV99 requires  fiducial sources of turbulence. The fluid in these sources, as most astrophysical fluids, has large Reynolds numbers. In fact,  $R_{e} = LV/\nu \sim 10^{20}$ (where V corresponds to a characteristic velocity of the fluid and $\nu$ is the kinematic viscosity  which for a magnetized fluid is dominated by transverse kinetic motions  to the magnetic field and is  given by $\nu \sim 1.7 \times 10^{-2}  n_c ln\Lambda/ (T^{0.5} B^2)\ \rm{cm^2 s^{−1}}$, being  $ln \Lambda \sim 25$  the Coulomb logarithm and $n_c$ is the coronal particle number density given by eq. 3 below). Likewise, the magnetic Reynolds number is $R_{em} = LV/\eta \sim10^{18}$ (where the magnetic diffusion coefficient 
 $\eta$  in the regime of strong magnetic fields is given by  $\eta = 1.3 \times 10^{13} cm{^2} s^{−1}  Z ln \Lambda T^{-3/2}$ (\citealt{spitzer62}).
As argued in KGL14, such  
high Reynolds numbers imply  that both the fluid and the magnetic field lines  can be highly distorted and turbulent if there is  turbulence triggering. In other words,  
any instability as for instance,  current driven instabilities, can naturally drive turbulence with characteristic velocities around the particles thermal speed.  Also, the occurrence of continuous magnetic reconnection during the building of the corona itself in the surrounds of the BH \citep{liu03} will contribute to the onset of turbulence which will then be further fed by fast reconnection as in LV99 model. Numerical simulations of coronal disk accretion also indicate the formation of turbulent flow in the surrounds of the BH that may be triggered by magnetorotational instability 
\citep[see e.g,][]{tchekhovskoy11, mckinney12, dexter_etal_14}. All these processes may ensure the presence of embedded turbulence in the magnetic discontinuity described in Figure 1.

We should also note that in the equations below which describe the accreting and coronal flow around the BH, we  adopt a nearly non relativistic approximation. In KGS14, we give quantitative arguments that indicate that this is a reasonable assumption. For instance, the evaluation of the magnetic reconnection power considering a pseudo-Newtonian gravitational potential to reproduce general relativistic effects,  gives a value that is similar to the classical case. A kinematic relativistic approach for the accreting and coronal flows is not necessary either since we are dealing with characteristic ion/electron temperatures smaller than or equal   $ \sim 10^9 \rm{K}$ (see KGS14).   Nevertheless, 
with regard to reconnection, the fact  that  $v_A$ may approach the light speed, may imply that relativistic effects can affect the turbulent driven fast reconnection. This question has been addressed in some detail in KGL14 as well, and we refer to this work (and the references therein). The current results indicate that one can treat both cases in a similar way. In particular, a recent study  \citep{cholazarian14} has demonstrated  that relativistic collisional MHD turbulence behaves as in the non-relativistic case which indicates that LV99 theory can be also applicable in the nearly relativistic regime.

Considering the assumptions above, KGS14 have demonstrated that the  magnetic power released by a fast magnetic reconnection event driven by turbulence in the corona around the BH, is given by:

\begin{equation}
W \simeq 1.66\times 10^{35} \Gamma^{-0.5} r_X^{-0.62} l^{-0.25} l_X q^{-2}\xi^{0.75}m\ ~ {\rm erg~s^{-1}},
\end{equation}
Where $r_X=R_X/R_S$ is the inner radius of the accretion disk in units of the BH Schwartzchild radius ($R_S$) (as in KGS14, in our calculations we assume $r_X=6$);  $l=L/{R_S}$ is the height of the corona in units of $R_S$; $l_X={L_X}/{R_S}$ where $L_X$ is the extension of the magnetic reconnection zone (as shown in Figure 1; see also Tables 1 and 2); $q=[1-(3/r_X)^{0.5}]^{0.25}$; $\xi$ is the  mass accretion disk rate in units of the Eddington rate ($\xi=\dot{M}/\dot{M}_{Edd}$)
which  we  assume to be  $\xi \simeq 0.7$\footnote{We note that according to the results of KGS14 (see their Figure 5), accretion rates $\xi$  between $0.05 < \xi \leq  1$ are able to produce magnetic reconnection power values which are large enough to probe the observed luminosities from microquasars. We here adopted $\xi\simeq 0.7$ as a fiducial value.}; $m$ is the BH mass in units of solar mass, and $v_A = v_{A0} \Gamma$, is the relativistic form of the Alfv\'en velocity,   with $v_{A0} = B/(4\pi \rho)^{1/2}$,  $B$ being the local magnetic field, $\rho$  the fluid density, and  $\Gamma=[1+({v_{A0} \over c})^2]^{-1/2}$  \citep{somov12}. In this  work,  $v_{A0} \sim c$ (see below).

The ambient magnetic field in the surrounds of the BH calculated from the GL05 and KGS14  model is given by:
\begin{equation}
B\cong 9.96\times 10^{8}r_X^{-1.25} \xi^{0.5}m^{-0.5}\ {\rm G}.
\end{equation}

The particle density  in the coronal region in the surrounds of the BH is 
\begin{equation}
n_c\cong 8.02\times 10^{18}r_X^{-0.375} \Gamma^{0.5} l^{-0.75}q^{-2}\xi^{0.25}m^{-1}\ {\rm cm^{-3}}.
\end{equation}

The equations above will be employed in Sections 4 and 5 to model the acceleration in the core region of the microquasars Cyg  X-1 and Cyg X-3.
The acceleration region in our model is taken to be the cylindrical shell where magnetic reconnection takes place, as in Figure 1. This shell has a length $l_X$, and inner and outer radii  $R_X$ and $R_X+\Delta R_X$ respectively, where $\Delta R_X$ is the width of the current sheet  given by (KGS14): 
\begin{equation}
\Delta R_X\cong 2.34\times 10^{4} \Gamma^{-0.31} r_X^{0.48} l^{-0.15} l_X q^{-0.75}\xi^{-0.15}m\ {\rm cm}.
\end{equation}

In sections 4 and 5, we will also need the accretion disk temperature in order to evaluate its black body radiation field:

\begin{equation}
T_d\cong 3.71\times 10^{7}\alpha^{-0.25}r_X^{-0.37}m^{0.25}\ \rm K,
\end{equation}
where $0.05 \leq \alpha < 1$ is the Shakura-Sunyaev disk viscosity parameter which we here assume to be of the order of 0.5.

\subsection{Particle acceleration due to the magnetic energy released by fast reconnection}

The magnetic power released by a fast reconnection event heats the surrounding gas and may accelerate particles. 
We assume that approximately $50\%$ of the reconnection power is used to accelerate the particles. This  is consistent with recent plasma laboratory experiments of particle acceleration in reconnection sheets (e.g., \citealt{yamada}) and also with solar flare observations where up to $50\%$ of the released magnetic energy appears in the form of energetic electrons (e.g., \citealt{lin71}).

As  in shock acceleration where particles confined between the upstream  and downstream flows undergo a first-order Fermi acceleration, GL05  proposed that a similar mechanism would occur when particles are trapped between the two converging magnetic flux tubes moving to each other  in a magnetic reconnection current sheet with a velocity $V_{R}$. They showed that, as particles bounce back and forth  due to head-on collisions with magnetic fluctuations in the current sheet, their energy after a round trip increases by $< \Delta E/E > \sim 8 V_{R}/3c$, which implies  a first-order Fermi process with an exponential energy growth after several round trips (GL05; see also \citealt{bete13}). 
Under conditions of $fast$ magnetic reconnection  $V_{R}$ is of the order of the local Alfv\'en speed $V_{A}$, at the surroundings of relativistic sources, $V_{R} \simeq  v_A \simeq c$ and thus  the mechanism can be rather efficient (GL05, \citealt{Giannios2010}).

As remarked earlier, this mechanism has been thoroughly tested by means of 3D MHD numerical simulations in which charged thermal particles are accelerated to relativistic energies into collisional domains of fast magnetic reconnection without including kinetic effects (\citealt{kowal2011,kowal2012}).\footnote{We note also that tests performed in collisionless fluids, by means of 2D (e.g. \citealt{zenitani01,zenitani09,drake06,drake10,cerutti13,cerutti14,sironi14}), and 3D PIC simulations (\citealt{sironi14}) have generally achieved similar results to those of collisional studies with regard to acceleration rates and particle power law spectra, with the only difference that these can probe only the kinetic scales of the process, while the collisional MHD simulations probe large scales (\citealt{kowal2011,kowal2012}). 
In particular, \cite{kowal2011} have demonstrated by means of 2D and 3D collisional MHD simulations the equivalence between first-order Fermi particle acceleration involving 2D converging magnetic islands in current sheets, which arise in collisionless fluid simulations (e.g., \citealt{drake06, drake10}), and the same process in 3D reconnection sites where the islands naturally break out into loops. \cite{kowal2011} further demonstrated the importance of the presence of guide fields in 2D simulations to ensure equivalence with the results of more realistic 3D particle acceleration simulations. }

Using the results of the 3D MHD numerical simulations of the acceleration of test  particles  in current sheets where reconnection was made fast by embedded turbulence \citep{kowal2012}, we find that the  acceleration rate for a proton is given by:
\begin{equation}
t^{-1}_{acc,rec,p}=1.3\times 10^5\left(\frac{E}{E_0}\right)^{-0.4}t_0^{-1},
\end{equation}
where $E$ is the energy of the accelerated proton, $E_0=m_p c^2$, $m_p$ is the proton rest mass, $t_0=l_{acc}/{v_A}$  is the Alfv\'en time,  and $l_{acc}$ is the length scale of the acceleration region. 
Although this result was found from numerical simulations employing protons as test particles, we can derive a similar expression for the electrons:
\begin{equation}
t^{-1}_{acc,rec,e}=1.3\times 10^5\sqrt{\frac{m_p}{m_e}}\left(\frac{E}{E_0}\right)^{-0.4}t_0^{-1},
\end{equation}
where $m_e$ is the electron rest mass.

The equations above will be used to compute the acceleration rates in our model as described  in the following sections.

The accelerated particles develop a  power law energy distribution (see also Appendix A): 
\begin{equation}
Q(E)\propto E^{-p},
\end{equation}
we assume for the power law index  $p=1.8$ and $p=2.2$ for Cyg X-1 and Cyg X-3, respectively, which are compatible with the predicted values in analytical and numerical studies (GL05, \citealt{Drury2012,bete13,delvalle14}).\footnote{We note that analytical estimates of the first-order Fermi accelerated particle power law spectrum in current sheets  predict power law indices $p \sim 1-2.5$ (e.g., GL05, \citealt{Drury2012,Giannios2010}), while 3D MHD numerical simulations with test particles predict $p \sim 1$ (\citealt{kowal2012} and see also the review by \citealt{bete13}),  which is  comparable with results obtained from 2D collisionless PIC simulations considering merging islands  $p \sim 1.5$ (\citealt{drake10}), or X-type Petschek 2D configurations  (e.g., Zenitani \& Hoshino 2001), for which it has been obtained $p\sim 1$, or even with  more recent 3D PIC simulations  (\citealt{sironi14}) which obtained $p < 2$. In summary, considering both analytical and numerical predictions  $p \sim 1-2.5$. However, at least in the case of the 3D MHD simulations,  some caution is necessary with the derived spectral index $p \sim 1$, because in these simulations  particles are allowed to re-enter in the periodic boundaries of  the computational domain and be further accelerated causing some deposition of particles in the very high energy tail of the spectrum after saturation of the acceleration which may induce some artificial increase in the slope \citep{delvalle14}.  }

As stressed in GL05,  it is also possible that a diffusive shock may develop in the surrounds of the magnetic reconnection zone, at the jet launching region,  due to the interaction of "coronal mass ejections", which are released by fast reconnection along the  magnetic field lines, just like, e.g.,  in the Sun. A similar picture has been also suggested by e.g., \cite{romerovieyro2010}. In this case,  one should expect the shock velocity  to  be predominantly parallel to the magnetic field lines and the  acceleration rate  for a particle of energy $E$ in a magnetic field $B$, will be approximately given by  (e.g.,  \citealt{spruit88}):
 \begin{equation}
 t^{-1}_{acc,shock}=\frac{\eta e c B}{E},
 \end{equation}
where $0<\eta \ll 1$ characterizes the efficiency of the acceleration. We fix $\eta=10^{-2}$, which is appropriate for shocks with velocity $v_s\approx 0.1c$  commonly assumed in the Bohm regime  (\citealt{romerovieyro2010}) (see further discussion in Section 6).

The accelerated particles lose their energy radiatively via interactions with the surrounding magnetic field (producing synchrotron emission), the photon field (producing inverse Compton, synchrotron-self-Compton, and photo-mesons $p\gamma$), and with the surrounding matter (producing $pp$ collisions and relativistic Bremsstrahlung radiation).

In the following section we discuss the relevant radiative loss processes for electrons and protons which will allow the construction of the SED of these sources for comparison with the observations.

\section{Emission and Absorption Mechanisms}

\subsection{Interactions with magnetic field}
Charged particles with energy $E$, mass $m$ and charge number $Z$ spiralling in a magnetic field $\vec{B}$ emit synchrotron radiation at a rate
\begin{equation}
t^{-1}_{synch}(E)=\frac{4}{3}\left(\frac{m_e}{m}\right)^3 \frac{\sigma_T B^2}{m_e c8\pi}\frac{E}{mc^2},
\end{equation}
where $m_e$ is the electron mass and $\sigma_T$ is the Thompson cross section. The synchrotron spectrum radiated by a distribution of particles $N(E)$ (see appendix A) as function of the scattered photon energy $(E_\gamma)$ (in units of power per unit area) is 
\begin{dmath}
L_\gamma(E_\gamma)=\frac{E_\gamma V_{vol}}{4\pi d^2}\frac{\sqrt{2}e^3 B}{hmc^2}\int_{E_{min}}^{E_{max}}dE N(E) \frac{E_\gamma}{E_c}\int_{\frac{E_\gamma}{E_c}}^\infty K_{5/3}(\xi)d\xi,
\end{dmath}
where $V_{vol}$ is the volume of the emission region, $d$ is the distance of the source from us,  $h$ is the Planck constant, $K_{5/3}(\xi)$ is the modified Bessel function of 5/3 order, and the characteristic energy $E_c$ is 
\begin{equation}
E_c=\frac{3}{4\pi}\frac{ehB}{mc}\left(\frac{E}{mc^2}\right)^2.
\end{equation}
In these calculations we assumed that the particle velocity is perpendicular to the local magnetic field.

 To compute  equation (11) we used the approximation
\begin{equation}
x\int_x^\infty K_5/3(\xi)d\xi\approx1.85x^{1/3}e^{-x}.
\end{equation}

Practically, the synchrotron emission of the electrons dominates the low energy photon background which is a proper target for both inverse Compton (IC) and $p\gamma$ interactions  (see below; see also \citealt{reynosoetal2011}). 
The number density for multi-wavelength synchrotron scattered photons (in units of energy per volume), has been approximated as (\citealt{zhang2008})
\begin{equation}
n_{synch}(\epsilon)=\frac{L_\gamma(\epsilon)}{\epsilon ^2V_{vol}}\frac{r}{c}4\pi d^2,
\end{equation}
where $r$ stands for the radius of the emission region and $\epsilon$ for the scattered synchrotron radiation energy. More precisely,  $\epsilon$ corresponds to the photon energy of the multi-wavelength target radiation field for SSC and $p\gamma$ interactions. The volume $V_{vol}$ of the emission region in our model is taken as the spherical region that encompasses the cylindrical shell where magnetic reconnection particle acceleration takes place in Figure 1. Considering that the cylinder extends up to $L$, then $r \simeq L$ and the effective emission zone in our model has an approximate volume $4 \pi L^3/3$ (see Tables 1 and 2). 

\subsection{Interactions with matter}

\subsubsection{Bremsstrahlung}
When a relativistic electron accelerates in the presence of the electrostatic field of a charged particle or a nucleus of charge $Ze$, Bremsstrahlung radiation is produced. For a fully ionized plasma with ion number density $n_i$, the Bremsstrahlung cooling rate is (\citealt{Berezinskii1990}):
\begin{equation}
t_{Br}^{-1}=4n_iZ^2r_0^2\alpha_f c\left[\ln \left(\frac{2E_e}{m_ec^2}\right)-\frac{1}{3}\right],
\end{equation}
where $r_0$ is the electron classical radius and $\alpha_f$ stands for the fine structure constant. The relativistic Bremsstrahlung spectrum (in units of power per unit area) is given by (\citealt{romeromaria2010})
\begin{equation}
L_\gamma(E_\gamma)=\frac{E_\gamma V_{vol}}{4\pi d^2}\int_{E_\gamma}^{\infty} n\sigma_B(E_e,E_\gamma)\frac{c}{4\pi}N_e(E_e)dE_e,
\end{equation} 
where
\begin{equation}
\sigma_B(E_e,E_\gamma)=\frac{4\alpha_f r_0^2}{E_\gamma}\Phi(E_e,E_\gamma),
\end{equation}
and
\begin{dmath}
\Phi(E_e,E_\gamma)=\left[1+\left(1-\frac{E_\gamma}{E_e}\right)^2-\frac{2}{3}\left(1-\frac{E_\gamma}{E_e}\right)\right]\times \left[\ln\frac{2E_e(E_e-E_\gamma)}{m_ec^2E_\gamma}-\frac{1}{2}\right].
\end{dmath}

\subsubsection{$pp$ interactions}
One relevant  gamma-ray production mechanism is the decay of neutral pions which can be created through inelastic collisions of the relativistic protons with nuclei of the corona that surrounds the accretion disk. In this case the cooling rate is given by (\citealt{kelner2006})
\begin{equation}
t^{-1}_{pp}=n_ic\sigma_{pp}k_{pp},
\end{equation}
where $k_{pp}$ is the total inelasticity of the process of value $\sim 0.5$. The corresponding cross section for inelastic $pp$ interactions $\sigma_{pp}$ can be approximately by (\citealt{kelner2009})
\begin{equation}
\sigma_{pp}(E_p)=\left(34.3+1.88L+0.25L^2\right)\left[1-\left(\frac{E_{th}}{E_p}\right)^4\right]^2 \rm mb,
\end{equation}
 where $\rm mb$  stands for milli-barn, $L=\ln \left(\frac{E_p}{1TeV}\right)$, and the proton threshold kinetic energy for neutral pion $(\pi^0)$ production is $E_{th}=2m_\pi c^2(1+\frac{m_\pi}{4m_p})\approx280$ MeV, where $m_\pi c^2=134.97$ MeV is the rest energy of  $\pi^0$ (\citealt{vilaaharonian2009}). This particle decays in two photons with a probability of 98.8\%.

The spectrum can be calculated by 
\begin{equation}
L_\gamma(E_\gamma)=\frac{E_\gamma ^2 V_{vol}}{4\pi d^2}q_\gamma(E_\gamma),
\end{equation}
where $q_\gamma (E_\gamma) \rm{(erg^{-1}cm^{-3}s^{-1})}$ is the gamma-ray emissivity.

For proton energies less than $0.1$ TeV, $q_\gamma(E_\gamma)$ is 
\begin{equation}
q_\gamma(E_\gamma)=2\int_{E_{min}}^\infty\frac{q_\pi(E_\pi)}{\sqrt{E_\pi^2-m_\pi^2c^4}}dE_\pi,
\end{equation}
where $E_{min}=E_\gamma+m_\pi^2c^4/4E_\gamma$ and $q_\pi(E_\pi)$ is the pion emissivity. An approximate expression for $q_\pi(E_\pi)$ can be calculated using the $\delta$-function (\citealt{aharonianatoyan2000}). For this purpose, a fraction $k_\pi$ of the kinetic energy of the proton $E_{kin}=E_p-m_p c^2$ is taken by the neutral pion(\citealt{vilaaharonian2009}). The neutral pion emissivity is then given by
\begin{dmath}
q_\pi(E_\pi)=cn_i\int\delta(E_\pi-k_\pi E_{kin})\sigma_{pp}(E_p)N_p(E_p)dE_p=\frac{cn_i}{k_\pi}\sigma_{pp}(m_pc^2+\frac{E_\pi}{k_\pi}) \, N_p(m_pc^2+\frac{E_\pi}{k_\pi}).
\end{dmath}

The target ambient nuclei density is given by $n_i$ and $N_p(E_p )$ stands for the energy distribution of the relativistic protons. 

For proton energies in the range GeV-TeV, $k_\pi\approx 0.17$ (\citealt{gaisser1990}), the total cross section $\sigma_{pp}(E_p)$ can be approximated by
\begin{dmath}
$$
\sigma_{pp}(E_p)\approx \left\{\begin{array}{rl}
30\left[0.95+0.06ln\left(\frac{E_{kin}}{1GeV}\right)\right] mb &\mbox{$E_{kin}\geq1 \rm GeV,$} \\0 &\mbox{$E_{kin}<1 \rm GeV.$}
\end{array} \right.
$$
\end{dmath}

For proton energies greater than 0.1 TeV, the gamma-ray emissivity is
\begin{dmath}
q_\gamma(E_\gamma)=cn_i\int_{E_\gamma}^\infty \sigma_{inel}(E_p)N_p(E_p)E_\gamma(\frac{E_\gamma}{E_p},E_p)\frac{dE_p}{E_p}=cn_i\int_0^1\sigma_{inel}(\frac{E_\gamma}{x})N_p(\frac{E_\gamma}{x})F_\gamma(x,\frac{E_\gamma}{x})\frac{dx}{x}.
\end{dmath}

The inelastic $pp$ cross section is approximately given by
\begin{equation}
\sigma_{inel}(E_p)=(34.3+1.88L+0.25L^2)[1-(\frac{E_{th}}{E_p})^4]^2    mb,
\end{equation}
 Here $E_{th}=m_p+2m_\pi+\frac{m_\pi^2}{2m_p}=1.22$ GeV is the threshold energy of the proton to produce neutral pions $\pi^0$ and the number of photons whose energies are in the range of $(x,x+dx)$ where $x=E_\gamma/E_p$, caused per $pp$ collision can be approximated by (\citealt{vilaaharonian2009})
\begin{dmath}
F_\gamma(x,E_p)=B_\gamma\frac{ln x}{x}[\frac{1-x^{\beta_\gamma}}{1+k_\gamma x^{\beta_\gamma}(1-x^{\beta_\gamma})}]^4\times[\frac{1}{lnx}-\frac{4\beta_\gamma x^{\beta_\gamma}}{1-x^{\beta_\gamma}}-\frac{4k_\gamma \beta_\gamma x^{\beta_\gamma} (1-2x^{\beta_\gamma})}{1+k_\gamma x^{\beta_\gamma}(1-x^{\beta_\gamma})}].
\end{dmath}

The best least-squares fits to the numerical calculations yield:
\begin{equation}
B_\gamma=1.30+0.14L+0.011L^2,
\end{equation}
\begin{equation}
\beta_\gamma=(1.79+0.11L+0.008L^2)^{-1},
\end{equation}
\begin{equation}
k_\gamma=(0.801+0.049L+0.014L^2)^{-1}.
\end{equation}
Where $L=ln(E_p/1TeV)$ and $0.001\leq x\leq 0.1$ (for more details see \citealt{vilaaharonian2009}).

\subsection{Interactions with the radiation field}

Energetic electrons transfer their energy to low energy  photons causing them to radiate at high energies (inverse Compton process). On the other hand, when high energy protons  interact with low energy photons ($p\gamma$ interactions) they produce pions and gamma-ray photons with energies larger than $10^8$ eV in the so called photomeson process.

\subsubsection{Inverse Compton}

The IC cooling rate for an electron in both Thomson and Klein-Nishina regimes is given by (\citealt{blumenthal70})
\begin{equation}
t_{IC}^{-1}(E_e)=\frac{1}{E_e}\int_{\epsilon_{min}}^{\epsilon_{max}}\int_{E_{ph}}^{\frac{{\Gamma E_e}}{1+\Gamma}}(E_\gamma-E_{ph})\frac{dN}{dt dE_\gamma}dE_\gamma.
\end{equation}
Here $E_{ph}$ and $E_\gamma$ are the incident and scattered photon energies, and
\begin{equation}
\frac{dN}{dtdE_\gamma}=\frac{2\pi r_0^2m_e^2c^5}{E_e^2}\frac{n_{ph}(E_{ph})dE_{ph}}{E_{ph}}F(q),
\end{equation}
where $n_{ph}(E_{ph})$ is the target photon density (in the units of $\rm{energy^{-1} volume^{-1}}$) and
\begin{equation}
F(q)=2q\ln q+(1+2q)(1-q)+0.5(1-q)\frac{(\Gamma q)^2}{1+\Gamma},
\end{equation}
\begin{equation}
\Gamma=4E_{ph}{E_e}/{(m_ec^2)^2},
\end{equation}
\begin{equation}
q=\frac{E_{\gamma}}{[\Gamma (E_e-E_\gamma)]}.
\end{equation}

Accelerated electrons may have interaction with photons produced by the synchrotron emission in the coronal region (eq. 14), in which case the process is SSC, or by photons emitted by the surface of the accretion disk.  This photon field can be represented by a  black body radiation and is given by
\footnote{We note that the contribution of target photons due to the radiation field produced by the companion star is found to be irrelevant in our model (e.g., \citealt{BoschRamonetal.(2005a)}).}
\begin{equation}
n_{bb}(E_{ph})=\frac{1}{\pi^2 \lambda_c^3m_ec^2}(\frac{E_{ph}}{m_ec^2})^2[\frac{1}{\exp({\frac{E_{ph}}{kt}})-1}].
\end{equation}
Here $\lambda_c$, $t$ and $k$ are the Compton wavelength, disk temperature and Boltzmann constant, respectively. We will see below that for the microquasars, the SSC will be dominating.

Taking into account the Klein-Nishina effect on the cross section, 
the total luminosity per unit area can be calculated from \citep{romeromaria2010}
\begin{dmath}
L_{IC}(E_\gamma)=\frac{E^2_\gamma V_{vol}}{4\pi d^2}\int_{E_{min}}^{E_{max}}dE_eN_e(E_e)\times\int_{E_{ph,min}}^{E_{ph,max}}dE_{ph} P_{IC}(E_\gamma,E_{ph},E_e),
\end{dmath}
where $P_{IC}(E_\gamma,E_{ph},E_e)$ is the spectrum of photons scattered by an electron of energy $E_e=\gamma_em_ec^2$ in a target radiation field of density $n_{ph}(E_{ph})$. According to \cite{blumenthal70}, it is given by
\begin{equation}
P_{IC}(E_\gamma,E_{ph},E_e)=\frac{3\sigma_tc(m_ec^2)^2}{4E^2_e}\frac{n_{ph}(E_{ph})}{E_{ph}}F(q),
\end{equation}
and for the scattered photons there is a range which is
\begin{equation}
E_{ph}\leq E_\gamma \leq \frac{\Gamma}{1+\Gamma}E_e.
\end{equation}

\subsubsection{Photomeson production $(p\gamma)$}
The photomeson production takes place for photon energies greater than $E_{th}\approx$ 145MeV. A single pion can be produced in an interaction near the threshold and then decay giving rise to gamma-rays. 
In our model the appropriate photons come from the synchrotron radiation.
\footnote{We find that  for  photomeson production, the  radiation from the accretion disk and from the companion star are irrelevant compared to the contribution from the synchrotron emission.\\} 
The cooling rate for this mechanism in an isotropic photon field with density $n_{ph}(E_{ph})$ can be calculated by \cite{Stecker1968}:
\begin{dmath}
t^{-1}_{p\gamma}(E_p)=\frac{c}{2\gamma_p^2}\int_{\frac{E_{th}^{(\pi)}}{2\gamma_p}}^\infty dE_{ph}\frac{n_{ph}(E_{ph})}{E^2_{ph}}\times \int_{E_{th}^{(\pi)}}^{2E_{ph}\gamma_p}d\epsilon_r \sigma_{p\gamma}^{(\pi)}(\epsilon_r)K_{p\gamma}^{(\pi)}(\epsilon_r)\epsilon_r,
\end{dmath}
where $\gamma_p=\frac{E_p}{m_ec^2}$,
 $\epsilon_r$ is the photon energy in the rest frame of the proton and $K_{p\gamma}^{(\pi)}$ is the inelasticity of the interaction. \cite{atoyandermer2003} proposed a simplified approach to calculate the cross-section and the inelasticity which are given by
\begin{equation}
$$
\sigma_{p\gamma}(\epsilon_r)\approx \left\{ \begin{array}{rl}
340\   \mu barn &\mbox{$300 {\rm MeV}\leq \epsilon_r\leq {\rm 500 MeV}$} \\
120\  \mu barn &\mbox{ $\epsilon_r>500 {\rm MeV},$}
\end{array} \right.
$$
\end{equation}
and
\begin{equation}
$$
K_{p\gamma}(\epsilon_r)\approx \left\{ \begin{array}{rl}
0.2\    &\mbox{$300 {\rm MeV}\leq \epsilon_r\leq 500 {\rm MeV}$} \\
0.6\   &\mbox{ $\epsilon_r>500 {\rm MeV}.$}
\end{array} \right.
$$
\end{equation}

To find the luminosity from the decay of pions, we use the analytical approach proposed by \cite{atoyandermer2003}. Taking into account that each pion decays into two photons, the $p\gamma$ luminosity is
\begin{dmath}
L_{p\gamma}(E_\gamma)=2 \frac{E^2_\gamma V_{vol}}{4\pi d^2}\int Q_{\pi^0}^{(p\gamma)}(E_\pi)\delta(E_\gamma-0.5E_\pi) dE_\pi=20 \frac{E^2_\gamma V_{vol}}{4\pi d^2}N_p(10E_\gamma)\omega_{p\gamma,\pi}(10E_\gamma)n_{\pi^0}(10E_\gamma),
\end{dmath}
where $Q_{\pi^0}^{(p\gamma)}$ is the emissivity of the neutral pions given by
\begin{equation}
Q_{\pi^0}^{(p\gamma)}=5N_p(5E_\pi)\omega_{p\gamma,\pi}(5E_\pi)n_{\pi^0}(5E_\pi),
\end{equation}
$\omega_{p\gamma}$ stands for the collision rate which is
\begin{dmath}
\omega_{p\gamma}(E_p)=\frac{m_p^2c^5}{2E_p^2}\int_{\frac{E_{th}}{2\gamma_p}}^\infty dE_{ph}\frac{n_{ph}(E_{ph})}{E_{ph}^2}\int_{E_{th}}^{2E_{ph}\gamma_p}dE_r
\sigma_{p\gamma}^{(\pi)}(E_r)E_r,
\end{dmath}
and $n_{\pi^0}$ is the mean number of neutral pions produced per collision given by
\begin{equation}
n_{\pi^0}(E_p)=1-P(E_p)\xi_{pn}.
\end{equation}

In the single-pion production channel, the probability for the conversion of a proton to a neutron with the emission of a $\pi^+-meson$ is given by $\xi_{pn}\approx0.5$. For photomeson interactions of a proton with energy $E_p$, the interaction probability is represented by $P(E_p)$, which is
\begin{equation}
P(E_p)=\frac{K_2-{\bar{K}}_{p\gamma}(E_p)}{K_2-K_1}.
\end{equation}

The inelasticity in the single-pion channel is approximated as $K_1\approx0.2$, whereas $K_2\approx0.6$. For energies above 500 MeV the mean inelasticity $\bar{K}_{p\gamma}$ is
\begin{equation}
\bar{K}_{p\gamma}=\frac{1}{t_{p\gamma}(\gamma_p)\omega_{p\gamma}(E_p)}.
\end{equation}

\subsection{Absorption}

Gamma-rays can be annihilated by the surrounding radiation field 
via electron-positron pair creation: $\gamma+\gamma \to e^++e^-$.  In microquasars, besides the radiation field of the tight companion star,  coronal and accretion disk photons can also absorb $\gamma-$rays. However, it has been  shown by \cite{cerutti11} that the absorption due to coronal photons  is negligible compared with the contribution from the disk.
Adopting the same absorption model for the disk radiation field of these authors we find that the disk contribution to gamma-ray absorption is less relevant  than that of the stellar companion, generally a Wolf-Rayet star, which produces  UV radiation.
To evaluate the optical depth due to this component, we have adopted the model described by \cite{sb08}, (see also  \citealt{dubus06,zdziarski}). 
This process is possible only above a kinematic energy threshold given by
\begin{equation}
E_\gamma \epsilon(1-\cos \theta)\geq 2m_e^2c^4,
\end{equation}
and
\begin{equation}
E_\gamma \epsilon>(m_ec^2)^2,
\end{equation}
in  head-on collisions (\citealt{romerovieyro2010}),
where $E_\gamma$ and $\epsilon$ are the energies of the emitted gamma-ray and the ambient photons and $\theta$ is the collision angle in the laboratory reference frame.
     
 The attenuated luminosity $L_\gamma(E_\gamma)$ after the $\gamma-$ray travels a distance $l$ is (\citealt{romero05})
 \begin{equation}
L_\gamma(E_\gamma)=L^0_\gamma(E_\gamma)e^{-\tau(l,E_\gamma)}
\end{equation}
 where $L^0_\gamma$ is the intrinsic coronal gamma-ray luminosity and $\tau(l,E_\gamma)$ is the optical depth.  The differential optical depth is given by:
 \begin{equation}
 d\tau=(1-\mu)n_{ph}\sigma_{\gamma \gamma}d\epsilon d\Omega dl'
 \end{equation}
 where $d\Omega$ is the solid angle of the target soft photons, $\mu$ is the cosine of the angle between the gamma-ray and the arriving soft photons, $l'$ is the path along the gamma-ray emission and $n_{ph}$ is the black-body photon density in $\rm cm^{-3}erg^{-1}sr^{-1}$.
 
 The $\gamma\gamma$ interaction cross-section $\sigma_{\gamma\gamma}$ is defined as (\citealt{gould67})
\begin{dmath}
\sigma_{\gamma\gamma}(\epsilon,E_\gamma)=\frac{\pi r_0^2}{2}(1-\beta^2)[2\beta(\beta^2-2)+(3-\beta^4)ln(\frac{1+\beta}{1-\beta})],
\end{dmath}
where $r_0$ is the classical radius of the electron and
\begin{equation}
\beta=[1-\frac{(m_ec^2)^2}{\epsilon E_\gamma}]^{1/2}.
\end{equation}

The companion star with radius $R_{\star}$ and  a black-body surface temperature $T_{\star}$ produces a photon density at a distance $d_{\star}$ from the star
\begin{equation}
n_{ph}=\frac{2\epsilon^2}{h^3c^3}\frac{1}{\exp(\epsilon/kT_{\star})}\frac{R_{star}^2}{d_{\star}^2}.
\end{equation}

In the absorption models proposed by \cite{sb08} and \cite{dubus06}, the geometrical parameters $d_{\star}$, $\mu$ and $l$ are strongly dependent on the viewing angle $\theta$ and the orbital phase $\phi_b$. In the superior conjunction, the compact object is behind the star and the orbital phase is  $\phi_b=0$. We here consider the same orbital phase that has been observed during the high energy observations for Cyg X-1 and Cyg X-3. (For more details on the geometrical conditions of the binary system and the integration extremes, see \citealt{sb08} and \citealt{dubus06}.)

We note that the  pairs produced by the absorbed gamma-rays  may  emit predominantly synchrotron emission in the surrounding magnetic fields (\citealt{bosch08}), but their emission is expected to be negligible compared to the other synchrotron processes of the system. We thus neglect this effect in our treatment of pair absorption (\citealt{zdziarski14}).


\section{Application to Cygnus X-1}
Cyg X-1 is a widely studied black hole binary system (\citealt{malyshev2013}) at a distance of 1.86-2.2 \rm{kpc} (\citealt{reid2011,ziolkowski2005}) which is accreting from a high mass companion star orbiting around the BH with a period is 5.6 days (\citealt{gies2008}).
The orbit inclination is between $25^{\circ}$ and $35^{\circ}$ (\citealt{gies1986}) with an eccentricity of $\sim$0.018 (\citealt{orosz11}), so that one can assume an approximate circular orbit with  a radius $r_{orb}$. 

The  parameters of the model for Cyg X-1 are tabulated in Table 1. The values for the first five parameters in the Table have been calculated from eqs. 1-5 above. We take for the accretion disk inner radius the value $R_X=6 R_S$, 
and for the extension $L_X$ of the reconnection region (see Fig. 1), we consider the value $L_X \simeq 10R_S$ (GL05, \citealt{beteluis2010a}).  
As remarked in Section 3, the volume $V$ of the emission region in Table 1 was calculated by considering the spherical region that encompasses the reconnection region in Figure 1.

The black hole mass has been taken from \cite{orosz11}. Figures 2 and  3 show the cooling rates for the different energy loss processes described in Section 3 (eqs. 10, 15, 19, 31 and 40) for electrons and protons. These are compared with the acceleration rates due to first-order Fermi acceleration by magnetic reconnection (eqs. 6 \& 7) and  to shock acceleration (eq. 9).

 \begin{table}
\centering
\begin{minipage}{82mm}
\caption{\label{arttype}Model parameters for Cyg X-1.}
\begin{tabular*}{\textwidth}{@{}llrrrrlrlr@{}}
\hline
$B$ &Magnetic field (\rm G)&$2.3\times10^7$\\
$n_c$&Coronal particle number density ($\rm cm ^{-3}$)&$4.5\times10^{16}$\\
$T_d$&Disk temperature (\rm K)&$4.4\times10^7$\\
$W$&Reconnection power (\rm erg/s)&$3.6\times10^{36}$\\
$\Delta R_X$&Width of the current sheet (\rm cm)&$1.1\times10^{7}$\\
$R_x$&Inner radius of disk (\rm cm)&$2.6\times10^7$\\
$L_X$&Height of reconnection region (\rm cm)&$4.3\times 10^7$\\
$V_{vol}$&Volume of emission region ($\rm cm^3$)&$3.5\times10^{23}$\\
$d$&Distance (\rm kpc)&$2$\\
$M$&Mass of BH ($M_{\odot}$)&$14.8$\\
$p$&Particle power index&$1.8$\\
$R_\star$&Stellar radius (cm)&$1.5\times10^{12}$\\
$T_\star$&Stellar temperature (K)&$3\times10^4$\\
$r_{orb}$&Orbital radius (cm)&$3.4\times10^{12}$\\
$\theta$&Viewing angle (rad)&$\pi/6$\\
\hline
\end{tabular*}
\end{minipage}
\end{table}

We notice that for both protons and electrons the acceleration is dominated by the first-order Fermi magnetic reconnection process in the core region. Besides, the main radiative cooling process for  the electrons is synchrotron radiation, while for protons the photo-meson production ($p\gamma$ interactions) governs the loss mechanisms (Figure 3). In this case,  the proper target radiation field are the photons from synchrotron emission. The intercept between the magnetic reconnection acceleration rate and the synchrotron rate in Figure 2 gives the maximum energy that the electrons can attain in this acceleration process, which is  $\sim10 \rm{GeV}$. 
Protons on the other hand, do not cool as efficiently as the electrons and can attain energies as high as $\sim4\times 10^{15}\rm{eV}$.

\begin{figure}
  \centering
  \includegraphics[width=0.45\textwidth]{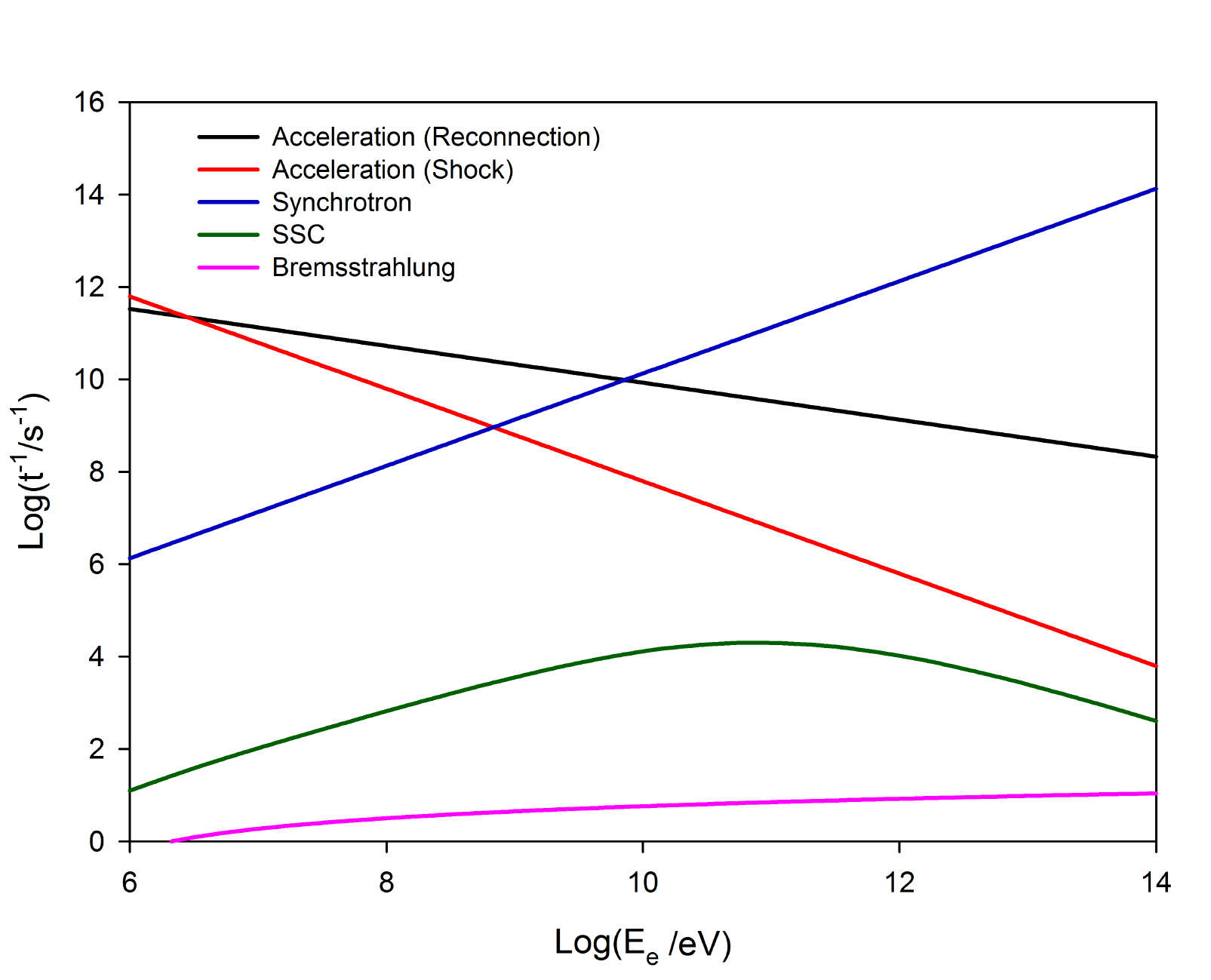}
  \caption{Acceleration and cooling rates for electrons in the nuclear region of Cyg X-1. }
  \label{wide_fig}
 \end{figure}
  
In order to reproduce the observed SED, we have calculated the non-thermal emission processes as described in Section 3 in the surrounds of the BH. Figure 4 shows the computed SED for Cyg X-1 compared with observed data. 
 As remarked, we have also considered the gamma-ray absorption due to electron-positron pair production resulting from interactions of the gamma-ray emission in the core with the surrounding radiation field.  As stressed, our calculations indicate that this process is dominated by  the  radiation field of the companion star. As a result, the opacity depends on the phase of the orbital motion  and  on the viewing angle.

The parameters employed in the evaluation of this absorption  are in the last four lines of Table 1, and have been taken from \cite{romeromaria2010}. It has been proposed from $MAGIC$ observations (\citealt{albert2007}) that the gamma-ray production and absorption are maximized near the superior conjunction (\citealt{b12}) at phase $\phi_b=0.91$. In our calculations we considered this orbital phase for Cyg X-1.

   The calculated opacity according to the equations above  results in a very high energy gamma ray absorption. We find that the produced gamma-rays are fully absorbed in the energy range of $50$ Gev-$0.5$ TeV which causes the energy gap seen in the  calculated SED in Figure 4. The observed upper limits by MAGIC plotted  in the diagram in this range are possibly originated outside the core,  along the jet where  $\gamma-$ray absorption by the stellar radiation is not important (see also \citealt{romeromaria2010}).
  \begin{figure}
  \centering
  \includegraphics[width=0.45\textwidth]{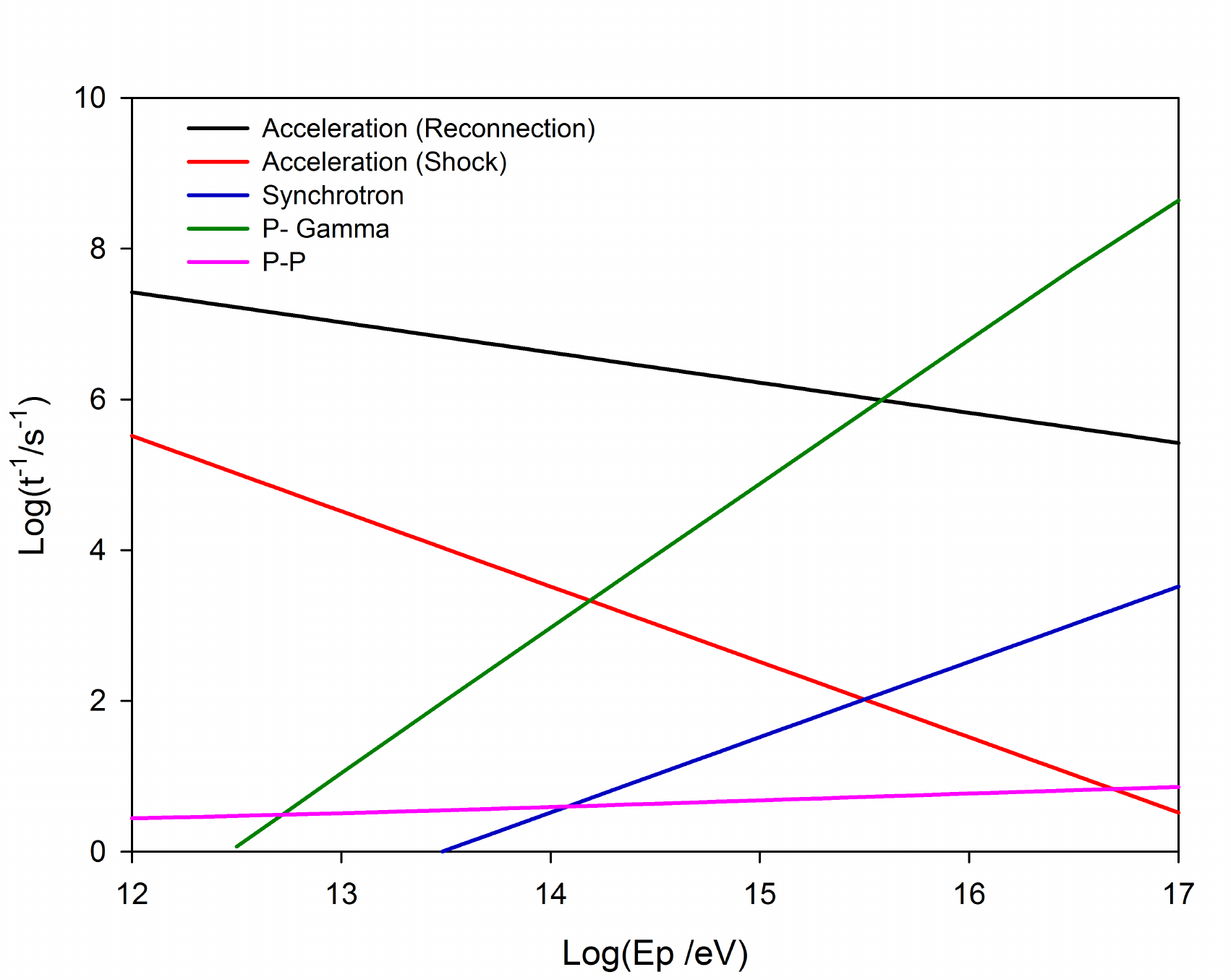}
  \caption{Acceleration and cooling rates for protons in the nuclear region of Cyg X-1.}
  \label{wide_fig}
 \end{figure}
 \begin{figure}
  \centering
  \includegraphics[width=0.50\textwidth]{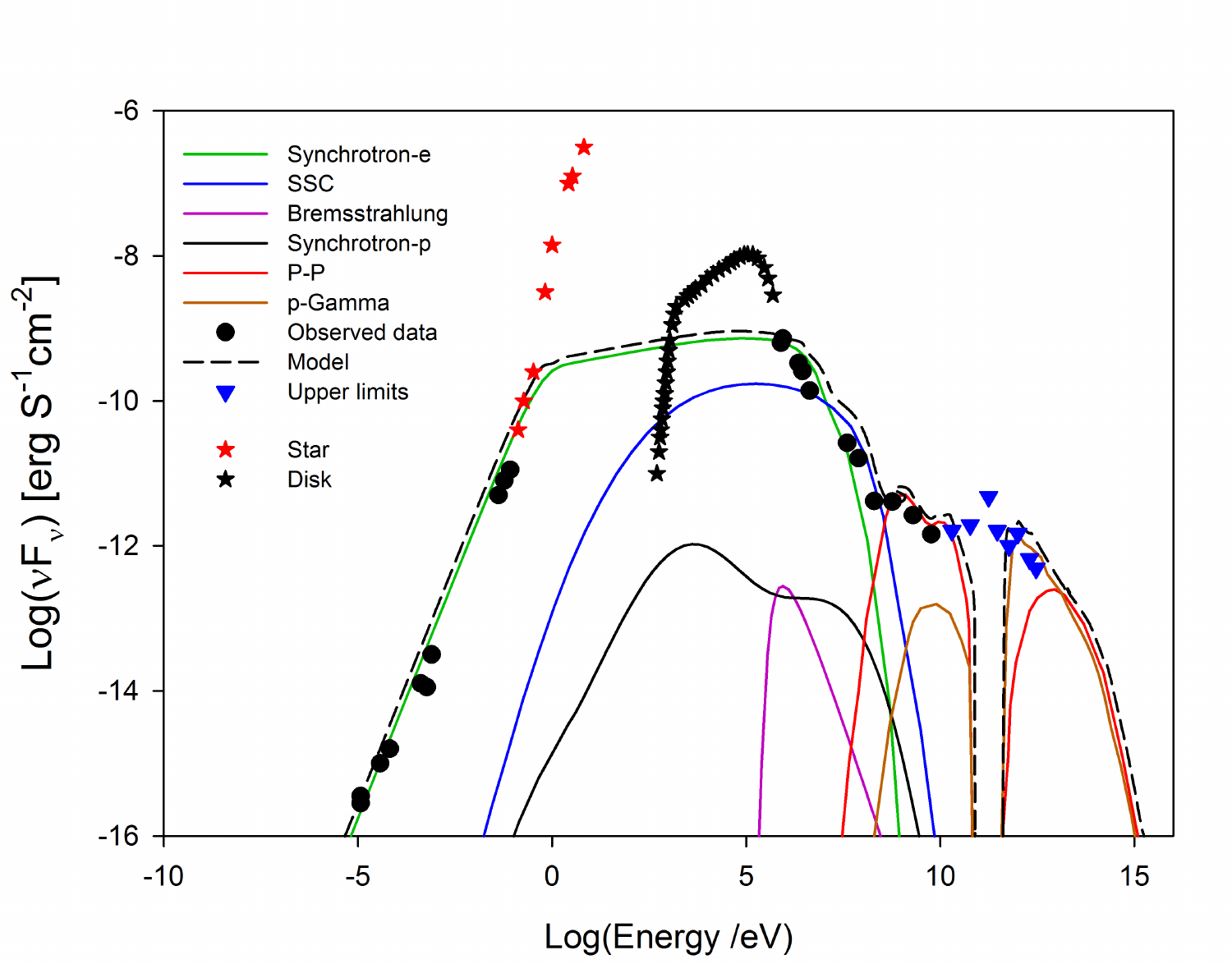}
  \caption{Calculated spectral energy distribution for Cyg X-1 using the magnetic reconnection acceleration model compared with observations. The data depicted in the radio  range is  from \citealt{fender00}, the IR fluxes are from \citealt{persi80,mirabel96}, the hard X-ray data above 20 $keV$  are from  INTEGRAL (\citealt{zdz12}), the soft X-ray  data below 20 $keV$  are from BeppoSAX (\citealt{disalvo01}),  the soft $\gamma$-ray data   are from COMPTEL (\citealt{mcconnel2000,mcconnel2002}), the data in the range 40 MeV- 40 GeV are measurements and upper limits from the Fermi LAT  (\citealt{malyshev2013}), and the data in the range 40 GeV- 3 TeV are  upper limits from MAGIC (with 95\% confidence level;  \citealt{albert2007}). The red and black stars correspond to emission from the companion star and the accretion disk, respectively, and are not investigated in the present model (see more details in the text.) }
  \label{wide_fig}
 \end{figure}
 
We note  that in Figure 4 the observed flux in  radio ($10$ $\mu\rm eV- 0.1$ eV) and  soft gamma-ray ($10^5-10^8$ eV) are explained by leptonic  synchrotron  and SSC processes according to the present model.  
In the range $10$ MeV- $0.2$ GeV,  SSC is the main mechanism to produce the observed data as a result of interactions between the high energy electrons with synchrotron photons.
 At energies in the range $0.2$ GeV - $3$ TeV, neutral pion decays reproduce the observed gamma-rays. These neutral pions result from $pp$  and $p\gamma$ interactions. 
In the range of $0.3$ GeV- $30$ GeV, $pp$ collisions are the dominant radiation mechanism, but in the very high energy gamma-rays,  interactions of relativistic hadrons (mostly protons) with scattered photons from synchrotron radiation  may produce the observed flux.

The observed emission in the near infrared ($0.1$ eV-$10$ eV), represented in Figure 4 by red stars is attributed to thermal blackbody radiation from the stellar companion, and the accretion X-ray emission ($1$ keV-$0.1$ MeV) 
also represented in Figure 4 by dark stars, is believed to be due to thermal Comptonization of the disk emission by the surrounding coronal plasma of temperature  $\sim 10^7$ K \citep{disalvo01,zdz12}. For this reason, these observed data are not fitted by the coronal non-thermal emission model investigated here. 

\section{Application to Cygnus X-3}
Cyg X-3 is also a high mass X-ray binary that possibly hosts a BH (\citealt{zdziarski}) and  a Wolf-Rayet as a companion star (\citealt{vankerkwijk}). The system is located at a distance of 7.2-9.3 kpc (\citealt{ling09}) and has an  orbital period of 4.8 h and an  orbital radius  $\approx3\times10^{11}\rm{cm}$ (\citealt{Pianoet2012}).
Our  model parameters  for Cyg X-3 are given in  Table 2. As in Cyg X-1,  the values for the first five parameters  were calculated from eqs. 1-5 which describe the magnetic reconnection acceleration model in the core region. 
We have also used for the accretion disk inner radius the value $R_X = 6 R_S$ and for the extension $L_X$ of the reconnection region the value $L_X=10 R_S$ (GL05, GPK10 and KGS14). The BH mass has been taken from \cite{sgs96}.

The cooling and acceleration rates for electrons and protons are depicted  in Figures 5 and 6, respectively. The maximum electron and proton energies in both diagrams are obtained from the intercept between the acceleration rate curve and the dominant radiative loss rate curve. As in Cyg X-1, it is clear from the diagrams that acceleration by magnetic reconnection is dominating over  shock acceleration in the core region.  Synchrotron emission is the main mechanism to cool the electrons which may reach energies as high as $\sim10 \rm{GeV}$, while the most important loss mechanism for protons is $p\gamma$ interactions with synchrotron photons. They can be accelerated up to $\sim 4\times 10^{15} \rm eV$. 

\begin{table}
\centering
\begin{minipage}{82mm}
\caption{\label{arttype}Model parameters for Cyg X-3.}
\begin{tabular*}{\textwidth}{@{}llrrrrlrlr@{}}
\hline
$B$ &Magnetic field (G)&$2.1\times10^7$\\
$n_c$&Coronal particle number density ($\rm cm ^{-3}$)&$3.9\times10^{16}$\\
$T_d$&Disk temperature (K)&$4.5\times10^7$\\
$W$&Reconnection power (erg/s)&$4.5\times10^{36}$\\
$\Delta R_X$&Width of the current sheet (\rm cm)&$1.3\times10^{7}$\\
$R_x$&Inner radius of disk (cm)&$3\times10^7$\\
$L_X$&Height of reconnection region (cm)&$5\times 10^7$\\
$V_{vol}$&Volume of emission region ($\rm cm^3$)&$5.3\times10^{23}$\\
$d$&Distance (kpc)&$8$\\
$M$&Mass of BH ($M_{\odot}$)&$17$\\
$p$&Particle power index&$2.2$\\
$R_\star$&Stellar radius (cm)&$2\times10^{11}$\\
$T_\star$&Stellar temperature (K)&$9\times10^4$\\
$r_{orb}$&Orbital radius (cm)&$4.5\times10^{11}$\\
$\theta$&Viewing angle (rad)&$\pi/6$\\
\hline
\end{tabular*}
\end{minipage}
\end{table}
 In this system, the close proximity $(R_d\approx3\times10^{11}\rm{cm})$, the large stellar surface temperature $(T_\star\sim10^5\rm{K})$, and the high stellar luminosity $(L_\star\sim10^{39}\rm{erg}$ $\rm{s^{-1}})$ of the Wolf$–$Rayet star may result a considerable  attenuation of the gamma-rays via $\gamma-\gamma$ pair production (\citealt{bednarek2010}). The detection of  TeV gamma-rays in Cyg X-3, therefore, relies on the competition between the production and the attenuation process above.

\begin{figure}
  \centering
  \includegraphics[width=0.47\textwidth]{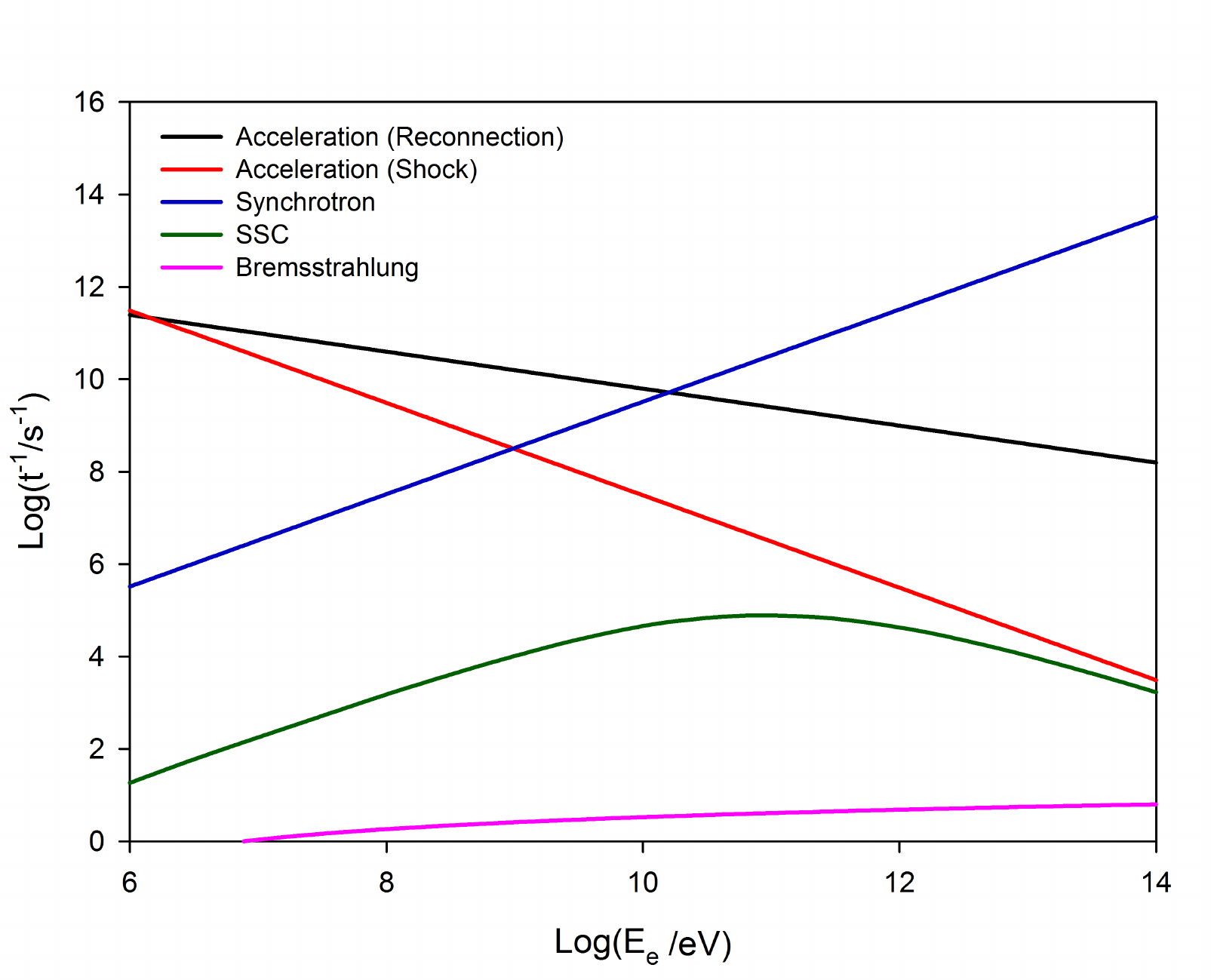}
  \caption{Acceleration and cooling rates for electrons in the core region of Cyg X-3.}
  \label{wide_fig}
 \end{figure}

Figure 7 shows the calculated SED compared to the observed data for this source.
The gamma-ray absorption was calculated from Eq. 52, employing  the UV field of the companion star which is a more significant target than the radiation fields of the accretion disk and the corona (see the stellar parameters in the last four lines of Table 2 which were taken from  \citealt{cm94}). The orbital phase considered was  $\phi_b=0.9$, near the superior conjunction (\citealt{aleksic10}), as in Cyg X-1.
The energy gap caused by this gamma-ray absorption is shown in Figure 7 in the  $50\rm{GeV}-0.4 \rm{TeV}$. 

  \begin{figure}
  \centering
  \includegraphics[width=0.47\textwidth]{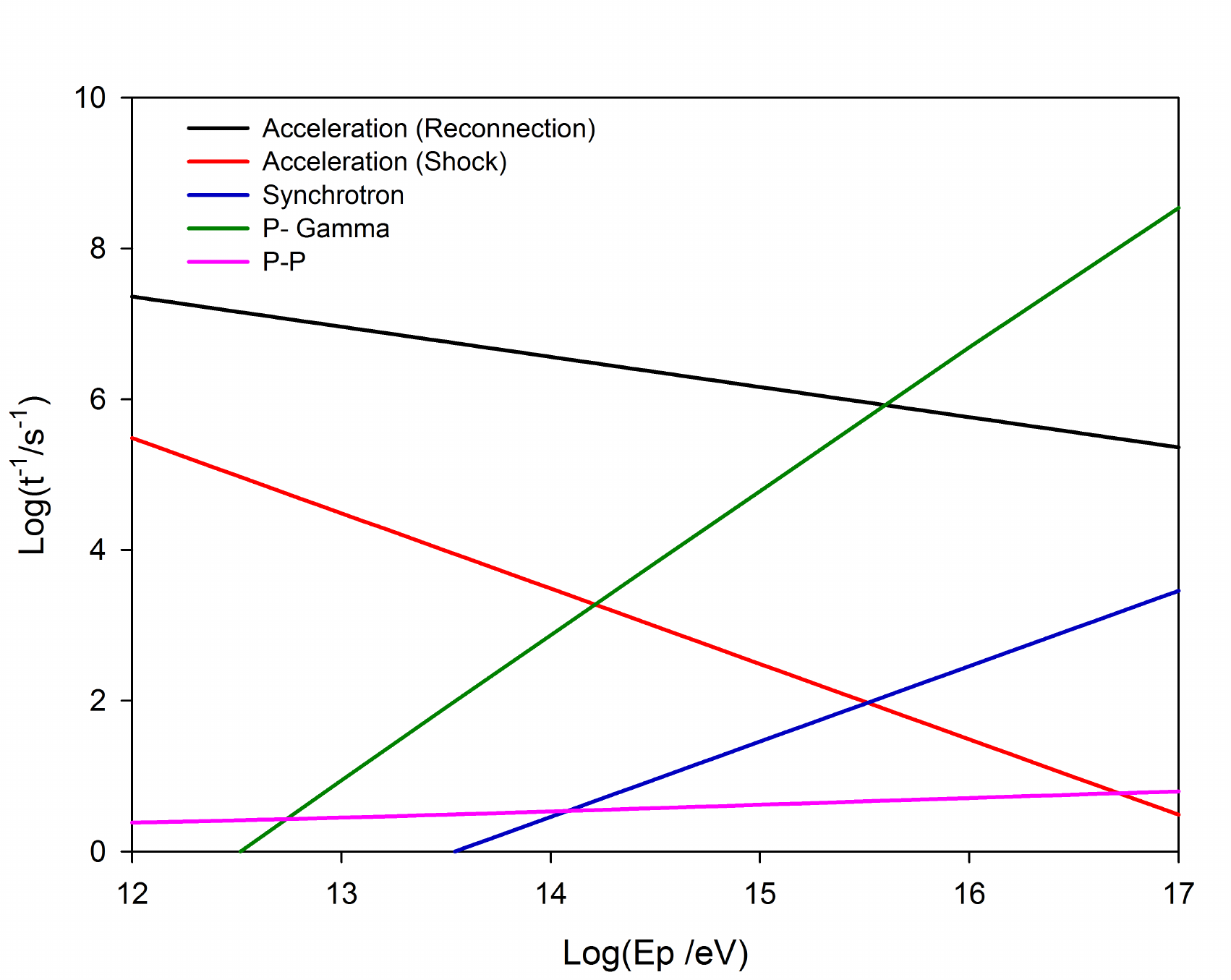}
  \caption{Acceleration and cooling rates for protons in the core region of Cyg X-3.}
  \label{wide_fig}
 \end{figure}
 \begin{figure}
  \centering
  \includegraphics[width=0.50\textwidth]{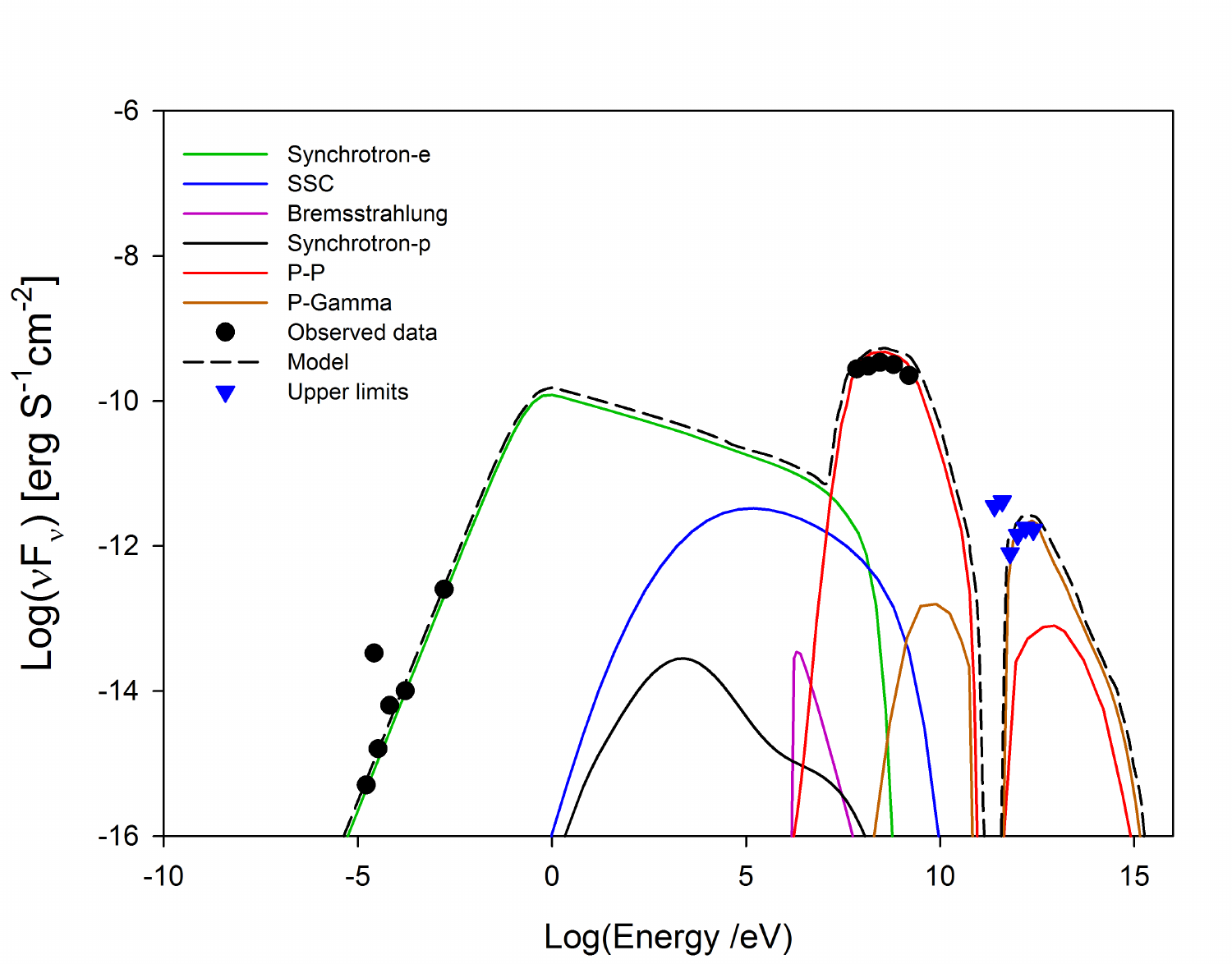}
  \caption{Spectral energy distribution for Cyg X-3. 
   The observed radio emission is taken from AMI-LA and RATAN \citep{Pianoet2012}; the data in the range $50 \rm MeV$ to $3 \rm GeV$ are from AGILE-GRID \citep{Pianoet2012}; and the data in the range $0.2-3.155 \rm TeV$ are  from MAGIC differential flux upper limits (95\% C.L.). }
  \label{wide_fig}
 \end{figure}
  The contributions of $pp$ and $p\gamma$ interactions are the dominant ones in the high energy gamma-ray range. These processes become more relevant in the coronal region around the BH since the magnetic field there is strong and enhances the synchrotron radiation of the electrons and protons. Also the matter and photon densities are large enough in the core region, providing dense targets for $pp$ and $p\gamma$ collisions and SSC scattering.
In the energy range $10 \rm MeV-50 \rm Gev$, the emission is dominated by the neutral pion decay resulting from $pp$ inelastic collisions. 
Also, the resulting interactions between accelerated protons and scattered photons from synchrotron emission  produce neutral pions and the gamma ray emission from these pion decays results in the tail seen in the SED for energies $\geq1\rm{TeV}$.

 \section{Discussion and Conclusions}
 The multi-wavelength detection, from radio to gamma-rays, of non-thermal energy from galactic black hole binaries (BHBs or $\mu QSRs$) is clear evidence of  an efficient production of relativistic particles and makes these sources excellent nearby laboratories to investigate and review particle acceleration theory in the surrounds of BH sources in general. 
Based on recent studies (GL05, GPK10, KGS14), we investigated here the role of magnetic reconnection in accelerating particles in the innermost regions of these sources, applying this acceleration model to reconstruct the spectral energy distribution (SED) of the BHBs Cyg X-1 and Cyg X-3.

According to GL05, particles can be accelerated to relativistic velocities in the surrounds of the BH, near the jet basis, by a first-order Fermi process occurring in the magnetic reconnection discontinuity formed by the encounter of  the magnetic field lines rising from the accretion disk with those anchored into the BH (Figure 1).  This process becomes very efficient when these two magnetic line fluxes are squeezed together by enhanced disk accretion and the reconnection is fast driven e.g., by turbulence \citep{LV99, kowal09, kowal2012}.

This driving mechanism  has been employed to compute the magnetic reconnection power released to heat and accelerate particles in this work (see KGS14). 
Moreover, the first-order Fermi acceleration mechanism within reconnection sites has been tested successfully by means of 2D and 3D  numerical simulations (e.g.,\citealt{kowal2011,kowal2012,drake06,zenitani09,sironi14}) and the resulting acceleration time scale is proportional to $\sim E^{0.4}$ (\citealt{kowal2012,delvalle14}). This can be compared with the typical estimated acceleration time scale  in diffusive shocks for the same environment conditions $t_{acc,shock}\propto E$ (see eq. 9). We find  a larger efficiency of the first mechanism  in regions where magnetic discontinuities are dominant.

It should be noted that in a  shock with perpendicular velocity to the magnetic field (for which particles diffuse $across$ the magnetic field lines), it is  predicted  that the acceleration rate may be larger than that resulting  from Bohm diffusion (Eq. 9) (\citealt{giacalone06, jokipii07,jokipii87, giacalone98, giacalone99}). As a matter of fact, if we consider the parameters in the inner coronal regions of our BHs, a
perpendicular  shock could lead to acceleration rates up to 2 or 3 orders of magnitude larger than that predicted by the Bohm rate, therefore,  comparable to the computed magnetic reconnection acceleration rates in Figures 2, 3, 5 and 6.  However, the model we explored here assumes that the surrounds of the BH is a magnetically dominated region, which makes the development of  strong shocks  harder in the inner nuclear regions. 
Nevertheless, as stressed in Section 3, fast magnetic reconnection can release coronal mass ejections along the reconnected magnetic field lines which will then induce the formation of a shock front further out, but in this case, the shock velocity will be predominantly parallel to the large scale magnetic field lines  and this explains why in Figures 2, 3, 5 and 6 we compared the magnetic reconnection acceleration rate with the Bohm shock acceleration rate which is suitable for diffusive and parallel shocks.  

Even if the presence of turbulence may allow the formation of important perpendicular magnetic field components in the shock location that may affect the shock acceleration rate, it is important to remark that recent results  (\citealt{lazarian14}) have demonstrated that the divergence of the magnetic field on scales less than the injection scale of the turbulence induces superdiffusion of cosmic rays (CRs) in the direction perpendicular to the mean magnetic field. This makes the square of the perpendicular displacement  to increase 
not  with the distance $x$ along the magnetic field, as in the case for a regular diffusion, but with $x^3$, for freely streaming CRs. They showed that this superdiffusion decreases the efficiency of the CR acceleration in perpendicular shocks. This superdiffusion has been also demonstrated numerically by \citealt{xu13} and these results  suggest that perpendicular shock acceleration efficiency is still an open question that deserves further  extensive numerical testing.  
A perpendicular shock would still be possible for  particular geometries of magnetic field lines 
as proposed by \citealt{jokipii87,giacalone99,giacalone06}, but this is out of the scope of the present work.

 As remarked earlier, fast magnetic reconnection has been detected in space environments, like the earth magnetotail and the solar corona (see e.g., \citealt{deng01,su13}). Striking evidence of turbulent reconnection in the flares and coronal events at work on the Sun have been provided by observations from the \textit{Yohkoh} and \textit{SOHO} satellites (\citealt{priest01}).
\cite{retino07} have also reported evidence of  reconnection in  the turbulent plasma of the solar wind downstream of the earth bow shock. They showed that this turbulent reconnection is fast and the released electromagnetic energy is converted into heating of the ambient plasma and  acceleration of particles. These findings have significant implications for particle acceleration within turbulent reconnection sheets not only in the solar, but also in astrophysical plasmas, in general.
Particle acceleration models due to fast magnetic reconnection have  been  widely explored in the solar framework. The Voyager spaceships completely failed to detect any observational evidence for shock acceleration. As the ultimate energy source in impulsive flares and in many other solar magnetic activities, fast reconnection naturally arose to explain the acceleration of  the observed anomalous cosmic rays throughout the heliosphere, from the solar flares and the earth magnetosphere (e.g., \citealt{drake06}) to the 
heliopause  (\citealt{lazarian09,drake10,oka10}).

Considering all the relevant  leptonic and hadronic radiative loss mechanisms due to the interactions of the accelerated particles with the surrounding matter, magnetic and  radiation fields in the core regions of the BHBs Cyg X-1 and Cyg X-3, we compared the time scales of these losses with the acceleration time scales above and found larger energy cut-offs  for particles being accelerated by magnetic reconnection than by a diffusive shock (see Figures 2 and 3 for Cyg X-1, and Figures 5 and 6 for Cyg X-3). These  cut-offs have  an important role in  the determination of the energy distribution of the accelerated particles and therefore, in the resulting SED, and stress the potential importance of magnetic reconnection as an acceleration mechanism in the core regions of BHBs and compact sources in general.

In most astrophysical systems, synchrotron is known as a dominant mechanism to cool the electrons and for the sources studied here, its cooling rate is also larger than that of the other loss mechanisms in all electron energy range. In Cyg X-1 and Cyg X-3, electrons gain energy up to 10 GeV (Figures 2 and 5). In  both cases, the achieved maximum energy are larger than the possible values obtained with Bohm-limit shock acceleration in the nuclear region.
Also, for both microquasars we find that  $p\gamma$ is the dominant mechanism to cool the accelerated relativistic protons in most of the investigated energy range.
Only for energies below $\sim 2$ TeV,  the $pp$ inelastic collisions are more efficient. The calculated energy cut-off for protons obtained from the comparison of the   $p\gamma$ cooling time with the magnetic reconnection acceleration time is $4\times 10^{15} \rm{eV}$, for both sources. In these $p\gamma$ processes, the synchrotron radiation is the dominant target photon field that interacts with the energetic protons, this because the magnetic field in the core region of these sources is relatively large, as calculated from Eq. 14.

We note that the maximum energy of the accelerated particles is not constrained only by the emission losses, but also by the size of the acceleration region, i.e., the particle Larmor radius, $r_L=E/ceB$, cannot be larger than the length scale of the acceleration zone. Considering the parameters employed in our model for both sources and $\Delta R_X$ as the length scale of the acceleration zone, we find that the maximum energy to which  the protons (and electrons) can be accelerated by magnetic reconnection is  $\sim 10^{17}\ \rm{eV}$, which is larger than the cut-off values obtained above. This  value also reassures the efficiency of this acceleration process. \

We have also shown  that, under fiducial  conditions, the acceleration model developed here is capable of explaining the multi-wavelength non-thermal SED  of both  microquasars Cyg X-1 and Cyg X-3. The radio emission may result from synchrotron process in both cases. 

The observed soft gamma-rays from Cyg X-1 are due to synchrotron and IC processes. The target photons for the IC come mainly from synchrotron emission (SSC). Neutral Pion decay resulting from $pp$ inelastic collisions may produce the high energy gamma rays in both systems, while the very high energy (VHE) gamma rays are the result of neutral pion decay due to photo-meson production ($p\gamma$) in the core of these sources.

 The importance of the $\gamma-\gamma$ absorption due to interactions with  the photon field of the  companion star for electron-positron pair production has been also addressed in our calculations. According to our results,  the observed gamma-ray emission in Cyg X-1 in the range  $5\times10^{10}-5\times10^{11} \rm{eV}$ (see inverted blue triangles in Figure 4) cannot be produced in  its core region  (see also \citealt{romeromaria2010}).  In the case of  Cyg X-3, we have  found  that the emission in the range of $50\rm{GeV}-0.4 \rm{TeV}$ (see inverted blue triangles in Figure 7) is also fully absorbed  in the core region by the same process. This suggests that in both sources, this emission is produced outside the core, probably along the jet, since  at larger  distances from the core the gamma ray absorption by the stellar companion decreases substantially. In fact, this is what was verified by   \cite{zhang14} in the case of  Cyg X-1.  
  
Other authors have proposed alternative scenarios to the one discussed here. The models  of \cite{Pianoet2012}, for instance, which were based on particle acceleration near  the compact object and  on  propagation along the jet, indicate that the observed gamma-ray $\leq$ 10 GeV in Cyg X-3 could be produced via leptonic (inverse Compton) and hadronic processes ($pp$ interactions). However, they have no quantitative estimates  for the origin of the VHE gamma-ray upper limits at  $\geq$0.1 TeV obtained by $MAGIC$.  \cite{sahakyan13}, on the other hand,  assumed that the jet of Cyg X-3 could accelerate both leptons and hadrons to high energies and the  accelerated protons escaping from the jet would  interact with the hadronic matter of the companion star producing  $\gamma-$rays and neutrinos. However, their model  does not provide  proper fitting in the TeV range either.

In the case of Cyg X-1,  \cite{zhang14} have employed a leptonic model to interpret recent Fermi LAT measurements also as due to  synchrotron emission but produced along the jet and to Comptonization of photons of the stellar companion. The TeV emission in their model  is attributed  to interactions between relativistic electrons and stellar photons via inverse Compton scattering. According to them this process could also explain the $MAGIC$ upper limits in the range of $50 \rm{GeV}-0.5 \rm{TeV}$, i.e., the band gap in Figure 4.  
 However, unlike the present work where we obtained a reasonable match due to $p\gamma$ interactions,  their model is unable to explain the observed upper limits by $MAGIC$ in the very high energy gamma-ray tail.

Also with regard to  Cyg X-1, we should note that the detection of  strong polarized signals in the high-energy  range of 0.4-2 MeV by \cite{laurent11} and \cite{jourdain12} suggests that the optically thin synchrotron emission of relativistic electrons from  the jet may produce  soft gamma-rays. There are indeed some theoretical models that explain the emission in this range by using a jet model \citep{zdz12,malyshev2013,zdziarski14,zhang14}. Nevertheless, contrary to this view,  \cite{romero14}  argue that the MeV polarized tail may  be originated in the coronal region of the core without requiring  the jet. This  study is therefore, consistent with the present model as it  supports the  coronal nuclear region for the origin of the non-thermal emission.

The results above clearly stress  the current uncertainties regarding the region where the HE and VHE emission are produced in these compact sources. This work has tried to shed some light on this debate focussing on a core model with a magnetically dominated environment surrounding the BH, but a definite answer to this question should be given by much higher resolution and sensitivity observations which may be achieved in near future with the forthcoming Cherenkov Telescope Array (CTA) \citep{actis11,acharya13,sol2013}.
  
We should also stress that there are two possible interpretations  for the lack of clear evidence of detectable TeV emission in Cyg X-1 and Cyg X-3. On one hand, there may be a strong absorption of these photons  by the ultraviolet (UV) radiation of the  companion  star (through the photon-photon process). On the other hand, the lack of emission may be  due to the limited time of observation \citep{sahakyan13}. 
In our model, we verified  that neutral pion decays due to $p\gamma$ interactions at  the emission region close enough to the central black hole, near the jet basis, could produce  TeV gamma-rays. Because of the high magnetic field near the black hole, a  large density synchrotron radiation field produced there could be a target photon field for the photo-meson production. These results predict  that a long enough observation time and higher sensitivity  would allow to capture substantial TeV $\gamma$-ray emission  from these microquasars. This may be also probed by the  CTA.

A final remark is in order. To derive the SEDs of the sources investigated here, we have assumed  a nearly steady-state accelerated particle energy distribution at the emission zone. This assumption is  valid as long as acceleration by  fast magnetic reconnection is sustained in the inner disk region, or in other words, as long as a large enough disk accretion rate is sustained in order to approach the magnetic field lines rising from  the accretion disk to those anchored into the BH. In microquasars, this should  last no longer than the time the system remains in the outburst state, normally ranging from less than one day to  several weeks.

\section*{Acknowledgements}
This work has been partially supported by grants from the Brazilian agencies FAPESP (2006/50654-3, and 2011/53275-4), CNPq (306598/2009-4) and CAPES. The authors are also indebted to Helene Sol for her useful comments on this work. EMGDP particularly thanks her kind hospitality during her visit to the Observatoire de Paris-Meudon.

\appendix 
\section{Particle energy distribution function}
The relativistic particles in the core region surrounding the BH may be accelerated up to  relativistic energies by a first-order Fermi process occurring within the  magnetic reconnection site. The injection and cooling of the accelerated particles occurs mainly in the coronal region around the black hole (see Figure 1).
We parametrize the isotropic injection function (in units of $\rm{erg^{-1} {\rm cm^{-3} s^{-1}}}$) as a power law with a high energy cut-off,
\begin{equation}
Q(E)=Q_0 E^{-p}exp{[-E/E_0]}
\end{equation}
with $p>0$ and $E_0$ is the cut-off energy. The normalization constant $Q_0$ is calculated from the total power injected in each type of particle
\begin{equation}
L_{(e,p)}=\int_{V_{vol}} d^3r \int_{E^{min}}^{E^{max}}dE\ E\ Q_{(e,p)}(E)
\end{equation}
where $L_{(e,p)}$  is the fraction of the magnetic reconnection power that accelerates the electrons and protons (see eq. 1 in the text).
The injection particle spectrum is modified in the emission region due to energy losses. We assume that  the minimum energy of the particles is given by $mc^2$, where $m$ is the rest mass of the particle\footnote{We note that the calculation of the emitted flux is little affected by the choice of the minimum energy of the particle spectrum.} 
and the maximum energy that the primary particles can attain is fixed by the balance of acceleration and the energy losses. Particles can gain energy up to a certain value $E_{max}$ for which the total cooling rate equals the acceleration rate.

The kinetic equation that describes the general evolution of the particle energy distribution $N(E, t)$ is the Fokker-Planck differential equation \citep{ginzburg64}.  We here use a simplified form of this equation. 
We employ the one-zone approximation to find the particle distribution, assuming that the acceleration region is spatially thin enough, so that we can ignore spatial derivatives in the transport equation. Physically, this means that we are neglecting the contributions to $N(E)$ coming from other regions than  the magnetic reconnection region in the inner accretion disk zone in the surrounds of the BH.
We consider a steady-state particle distribution which can be obtained by setting $\frac{\partial N}{\partial t}=0$ in the Fokker-Planck differential equation, so that the particle distribution equation is
\begin{equation}
N(E)=|\frac{dE}{dt}|^{-1} \int_E^\infty Q(E)dE.
\end{equation}
Here $-\frac{dE}{dt}\equiv E t_{cool}^{-1}$. It is very interesting to note that if the energy losses are proportional to the particle energy $(\frac{dE}{dt}\propto E), N(E)$ does not change the injection spectrum and $N(E)\propto E^{-p}$, as in the $pp$ inelastic collisions or Bremsstrahlung cool processes. In such loss mechanisms like synchrotron and IC scattering, in the Thomson regime, the $N(E)$ is steeper because in these cases $\frac{dE}{dt}\propto E^2$ and $N(E)\propto E^{-(p +1)}$.

The spectrum would be harder if $dE/dt$ were constant as for ionization losses, $N(E)\propto E^{-(p -1)}$. In the case of IC scattering in the Klein-Nishina limit, $\frac{dE}{dt}\propto E^{-1}$ and so, the spectrum is even harder and $N(E)\propto E^{-(p-2)}$.

\bsp

\label{lastpage}

\end{document}